\documentclass{article}

\makeatletter % `@' now normal "letter"
\@addtoreset{equation}{section}
\makeatother  % `@' is restored as "non-letter"

\usepackage{indentfirst}
\usepackage{setspace}
\usepackage{amsmath, mathtools}
\usepackage{mathrsfs}
\usepackage[colorlinks, linkcolor=black, anchorcolor=black, citecolor=black]{hyperref}
\usepackage{graphicx}
\usepackage{tocloft}
\title{\textbf{{{Effects of dark energy on the efficiency of charged AdS black holes as heat engines}}}}
\author{Hang Liu$^{a}$\thanks{E-mails:hangliu@mail.nankai.edu.cn} and Xin-He Meng$^{a,b}$\thanks{E-mails:xhm@nankai.edu.cn}
\\
\\
$^{a}$ \itshape School of Physics, Nankai University, Tianjin 300071, China\\
$^{b}$ \itshape State Key Laboratory of Theoretical Physics, Institute of Theoretical Physics,\\
\itshape Chinese Academy Of Science, Beijing 100190,China}
\date{}
\begin{document}
\large
\maketitle

\begin{abstract}
In this paper, we study the heat engine where charged AdS black holes surrounded by dark energy is the working substance and the mechanical work is done via $PdV$ term in the first law of black hole thermodynamics in the extended phase space. We first investigate the effects of a kind of dark energy (quintessence field in this paper) on the efficiency of the RN-AdS black holes as heat engine defined as a rectangle closed path in the $P-V$ plane. We get the exact efficiency formula and find that quintessence field can improve the heat engine efficiency which will increase as the field density $\rho_q$ grows. At some fixed parameters, we find that  bigger volume difference between the
smaller black holes($V_1$) and the bigger black holes($V_2$ ) will lead to a lower efficiency, while the bigger pressure difference $P_1-P_4$ will make the efficiency higher but  it is always smaller than 1 and will never be beyond Carnot efficiency which is the maximum value of the efficiency constrained by thermodynamics laws, this is consistent to the heat engine in traditional thermodynamics. After making some special choices for thermodynamical quantities, we find that the increase of electric charge $Q$ and normalization factor $a$ can also promote heat engine efficiency which would infinitely approach the Carnot limit when $Q$ or $a$ goes to infinity.

\end{abstract}

\setcounter{page}{0}
\tableofcontents\thispagestyle{empty}
\newpage
\section{Introduction}
Since Hawking and Page \cite{Hawking1} find that there exists a phase transition in the phase space  between  Schwarzschild-AdS black holes and thermal radiation(i.e. Hawking-Page phase transition), a lot of attentions have been paid to the study of the thermodynamical properties of AdS black holes \cite{Kubiznak1,Xu1,Cai1,Banerjee1,Banerjee2,Liu1,Gunasekaran1,Hendi1,Zou1,Zou2,Mirza1,Kastor1,Dolan1,Dolan2,Dolan3,Cvetic1,Belhaj1,
Hendi2,Spallucci1,Altamirano1,Hendi3,Hendi4,Hendi5,Hendi6}, especially to the study of AdS black hole thermodynamics in extended phase space in which the cosmological constant $\Lambda$ is not only treated as thermodynamical pressure $P$ by relation
\begin{equation}
P=-\frac{\Lambda}{8\pi}\label{4}
\end{equation}
but also a thermodynamical variable appearing in the first law of black hole thermodynamics such that the Smarr relation can be obtained. The $P-V$ criticality of RN-AdS black holes in extended phase space has been investigated by authors in Ref.\cite{Kubiznak1} with analogy between the AdS black holes and  van der Waals liquid-gas system, critical point and phase transition between large black hole and small black hole  are discovered in the refernce. Their study also showed that the electric charge $Q$ plays a crucial role in the critical behavior of AdS black holes, and the charged AdS black holes, as a gravity system, is highly similar to van der Waals liquid-gas system from thermodynamical perspective, and the two systems even share the same critical exponents. All of these discoveries suggest that the gravity system has a profound relation with the traditional thermodynamical system while people used to think that this two systems do not have any relations.

Inspired by the progress of the understanding of black hole thermodynamics, one natural idea is to introduce the concept of traditional heat engine into the black hole thermodynamics with both the thermodynamic pressure and volume defined in extended phase space. Johnson proposed this thought and firstly studied the charged AdS black holes as heat engine defined by a closed path in the $P-V$ plane in Ref.\cite{Johnson1} and calculated the efficiency of such heat engine. This amazing proposal implies  that the useful mechanical work of both static and stationary AdS black holes can be extracted from heat energy while the Penrose Process can only be used to  extract energy from rotating black holes, but Penrose Process can be exerted to black holes in both asymptotically AdS and flat spacetime. What's more, it is argued that such heat engine may have interesting holographic implications because the engine cycle represents a journal through a family of holographically dual large $\mathcal N$ field theories \cite{Johnson1}. Johnson's pioneering work has drawn lots of attentions and been generalized to many other black hole cases \cite{Mo1,Johnson2,Johnson3,Belhaj2,Setare1,Caceres1,Johnson4,Wei1,Zhang1,Sadeghi1,Sadeghi2,Bhamidipati1,Hennigar1,Hendi7,Mo2}.

\begin{figure}[thbp]
\centering
\includegraphics[height=3.5in,width=4.8in]{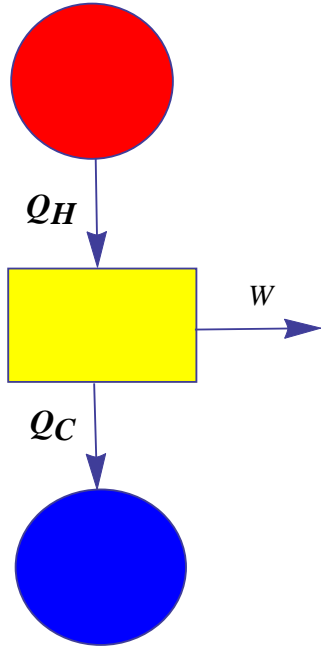}
\caption{The heat engine flows\label{fig1}}
\end{figure}

\begin{figure}[thbp]
\centering
\includegraphics[height=3in,width=4.3in]{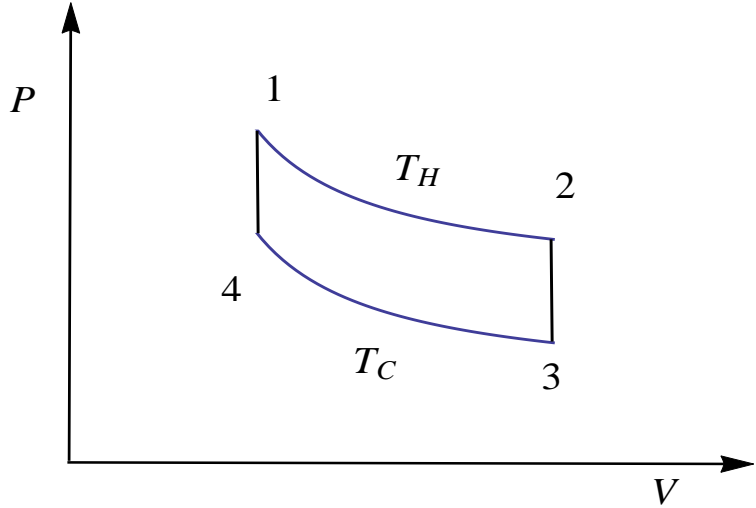}
\caption{Carnot engine\label{fig2}}
\end{figure}

We would like to briefly introduce Johnson's work \cite{Johnson1} in the following context. The figure \ref{fig1} represents the heat flows of the heat engine. The $Q_H$ represents the amount of input heat and $Q_C$ stands for the amount of exhaust heat, the work done by the heat engine is denoted as $W$. Obviously, the heat engine efficiency can be computed as
\begin{equation}
\eta=\frac{W}{Q_H}=\frac{Q_H-Q_C}{Q_H}=1-\frac{Q_C}{Q_H}
\end{equation}
For Carnot engine, the efficiency is
\begin{equation}
\eta_c=1-\frac{Q_C}{Q_H}=1-\frac{T_C}{T_H}\label{1}
\end{equation}
where $T_C$ is the lower temperature and $T_H$ is the higher temperature of the two heat sources which propel the heat engine to work. It is clear to see that the efficiency of Carnot engine is only dependent of the two temperatures of the heat sources. In figure \ref{fig2}, we give a kind of heat cycle of heat engine of static black hole and we show that this heat engine owns the same efficiency formula as Carnot heat engine. Along the upper isotherm (1$\rightarrow$ 2) of Fig \ref{fig2}, the amount of input heat is
\begin{equation}
Q_H=T_H\Delta S_{1\rightarrow2}=T_H\left(\frac{3}{4\pi}\right)^{\frac{2}{3}}\pi(V_2^{\frac{2}{3}}-V_1^{\frac{2}{3}})
\end{equation}
and along the lower isotherm (3$\rightarrow$4), we can get $Q_C$
\begin{equation}
Q_C=T_C\Delta S_{3\rightarrow4}=T_C\left(\frac{3}{4\pi}\right)^{\frac{2}{3}}\pi(V_3^{\frac{2}{3}}-V_4^{\frac{2}{3}})
\end{equation}
Note that we have $V_2=V_3,V_1=V_4$ which yields
\begin{equation}
\eta_c=1-\frac{Q_C}{Q_H}=1-\frac{T_C}{T_H}
\end{equation}
which is exactly the Carnot engine heat efficiency in Eq. (\ref{1}). The kernel of the calculations carried above is that the black hole entropy and thermodynamic volume are not independent and they both are the function of black hole horizon radius $r_h$ such that adiabats   coincide with isochores. It is worth noting that the example calculation presented above holds for all the static AdS black holes, in this sense, this calculation is a general consideration for the calculation of the efficiency of heat engines generated by static AdS black holes.

\begin{figure}[thbp]
\centering
\includegraphics[height=3in,width=4.5in]{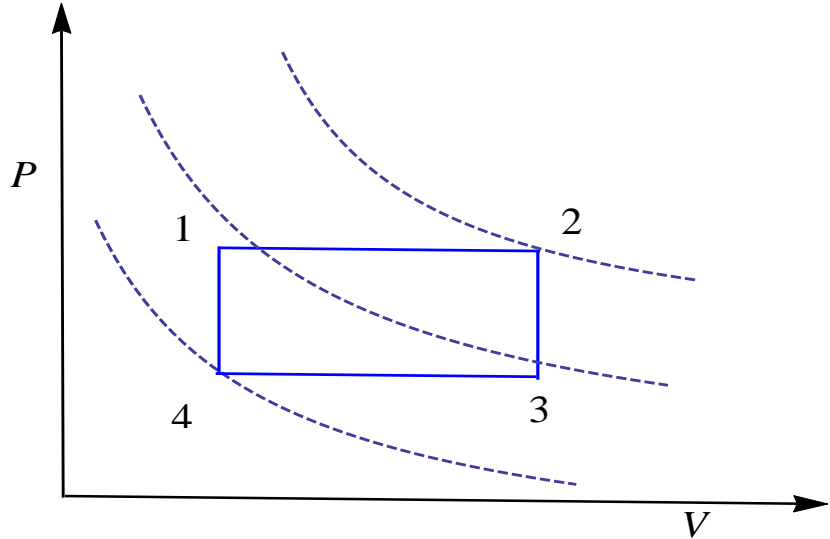}
\caption{our engine to be mainly discussed in this paper, and the dashed line represents the isothermals.\label{fig3}}
\end{figure}

Following the discussions above, we give an example \cite{Johnson1} of the efficiency calculations of RN-AdS black holes as heat engine defined in figure \ref{fig3} where the closed journey (1$\rightarrow$2$\rightarrow$3$\rightarrow$4$\rightarrow$1) is a rectangle such that it is very easy to calculate the work(i.e. the are of the rectangle) done by heat engine. The Hawking temperature on event horizon of RN-AdS black holes is
\begin{equation}
T=\frac{3r_+^4+l^2r_+^2-q^2 l^2}{4\pi l^2r_+^3}
\end{equation}
the cosmological radius $\ell$is related to cosmological constant in 4 dimension spacetime by relation $\Lambda=-3/\ell^2$.
Pressure $P$ can be expressed as
\begin{equation}
P=\frac{1}{8\pi}\left(\frac{4\pi}{3}\right)^{\frac{4}{3}}
\left(\frac{3T}{V^{\frac{1}{3}}}-\left(\frac{3}{4\pi}\right)^{\frac{2}{3}}\frac{1}{V^{\frac{2}{3}}}+\frac{q^2}{V^{\frac{4}{3}}}\right)
\end{equation}
heat capacity is
\begin{equation}
C=T\frac{\partial S}{\partial T}=\left(1-\frac{2S}{\sqrt \pi}\frac{\partial P}{\partial T}\right)
2S\left(\frac{8PS^2+S-\pi q^2}{8PS^2-S+3\pi q^2}\right)\label{3}
\end{equation}
where $S$ represents the black hole entropy. Heat capacity $C_P$ under constant pressure can be obtained by condition $\partial P/\partial T=0$. Note that we have
\begin{equation}
\left(\frac{\partial P}{\partial T}\right)_{V}=\frac{\sqrt \pi}{2\sqrt S}\label{2}
\end{equation}
we substitute Eq.(\ref{2}) into Eq.(\ref{3}) to calculate heat capacity $C_V$ under the constant volume, what we get is
$C_V=0$ implying that there is no heat exchange in the isochore process which means that this process is adiabatic. $Q_H$ is
\begin{equation}
Q_H=\int_{T_1}^{T_2}C_P(P,T)dT
\end{equation}
the output work $W$
\begin{equation}
W=\frac{4}{3\sqrt \pi}\left(S_2^{\frac{3}{2}}-S_1^{\frac{3}{2}}\right)(P_1-P_4)
\end{equation}
Expand entropy and $C_P$ at large $T$ and $P$
\begin{align}
S&=\frac{\pi}{4}\frac{T^2}{P^2}-\frac{1}{4P}-\frac{1}{16\pi T^2}+\cdots\\
C_P&=\frac{\pi}{2P^2}T^2+\frac{1}{8\pi T^2}+\cdots
\end{align}
to obtain approximate efficiency as
\begin{equation}
\eta=\frac{W}{Q_H}=(1-\frac{P_4}{P_1})\left\{1-\frac{3}{8P_1}\left(\frac{S_2^{\frac{1}{2}}-S_1^{\frac{1}{2}}}
{S_2^{\frac{3}{2}}-S_1^{\frac{3}{2}}}\right)+O\left(\frac{1}{P_1^2}\right)\right\}
\end{equation}
Actually, we have
\begin{equation}
Q_H=\int_{T_1}^{T_2}C_P(P,T)dT=\int_{T_1}^{T_2} T\frac{\partial S}{\partial T}\bigg |_{P}dT=\int_1^{2}TdS
\end{equation}
Considering the first law of thermodynamics under the constant pressure and other parameters except entropy
\begin{equation}
dM=TdS
\end{equation}
which yields
\begin{gather}
Q_H=M_2-M_1\\
Q_C=M_3-M_4
\end{gather}
so finally we have the exact efficiency
\begin{equation}
\eta=1-\frac{M_3-M_4}{M_2-M_1}\label{14}
\end{equation}

Modern observational results \cite{Bachall1,Perlmutter1,Sahni1} have showed that our universe is expanding with acceleration. One method to explain this amazing accelerating expansion is to introduce the so called dark energy which makes up about 70 percent of the universe and has negative pressure to propel the accelerating expansion of the cosmos.
Cosmological constant and scalar field are two kinds of dark energy model. Quintessence is proposed as spatially homogeneous canonical real scalar field believed as a type of candidate of dark energy  model with the state parameter $-1<\omega_q<1$, but the parameter must be restricted to $-1<\omega_q<-\frac{1}{3}$ to explain the late-time cosmic acceleration \cite{Fujii1,Ford1,Ratra1,Wetterich1}. Kiselev \cite{Kiselev1} obtained the exact solution of Einstein equation with the quintessence matter surrounding a  charged or non-charged black hole. And then, based on the Kiselev's achievement, considerable attentions \cite{Chen1,Chen2,Varghese1,Tharanath1,Ainou1,Wei2,Saleh1,Zeng2,Zeng1,Chen3} have been paid to investigate the thermodynamical properties of black holes surrounded by quintessence, including the influence of quintessence to the critical behavior in extended phase space of such black holes \cite{Li1}. Taking into account the dark energy, it would be of interest to investigate whether and how the  dark energy would influence the heat engine efficiency of AdS black holes and we believe that this may not only deep our understanding of black hole thermodynamics but also provide us a new hint to research dark energy .

This paper is organized as follows. In section 2, we would like to give a brief review of RN-AdS black holes surrounded by quintessence since it is the base of our present work. In section 3, we discuss the influence of quintessence to the heat engine efficiency of charged AdS black holes. In section 4, we will investigate critical black holes as heat engines and the last section  is devoted to conclusions and discussions.

\section{A brief review of thermodynamics of black holes surrounded by quintessence}
In this section, we are going to introduce the thermodynamics of RN-AdS black holes surrounded by quintessence by reviewing Ref.\cite{Li1,Kiselev1}.

The metric of such black holes have the form
 \begin{equation}
ds^2=f(r)dt^2-f(r)^{-1}dr^2-r^2(d\theta^2+\sin^2\theta d\varphi^2)\label{metric}
\end{equation}
where $f(r)$ is
\begin{equation}
f(r)=1-\frac{2M}{r}+\frac{Q^2}{r^2}-\frac{a}{r^{3\omega_q+1}}-\frac{\Lambda r^2}{3}
\end{equation}
the parameter $\omega_q$ and $a$ are the state parameter and normalization factor of quintessence field, respectively.
$r_+$ stands for the event horizon radius meeting the horizon equation $f(r_+)=0$ which yields the black hole mass $M$
\begin{equation}
M=\frac{r_+}{2}\left(1+\frac{Q^2}{r_+^2}-\frac{a}{r_+^{3\omega_q+1}}-\frac{\Lambda r_+^2}{3}\right)
\end{equation}
Hawking temperature related to surface gravity $k$ on event horizon is
\begin{equation}
T=\frac{k}{2\pi}
\end{equation}
where $k$ can be calculated by the general formula
\begin{equation}
k=\lim_{r\rightarrow r_+}\left[-\frac{1}{2}\sqrt{\frac{g^{11}}{-g_{00}}}\frac{\partial g_{00}}{\partial r}\right]
\end{equation}
and then hawking temperature can be computed as
\begin{equation}
T=\frac{f'(r)}{4\pi}=\frac{1}{4\pi}\left(\frac{1}{r_+}-\frac{Q^2}{r_+^3}+\frac{3a\omega_q}{r_+^{2+3\omega_q}}-r_+\Lambda\right)\label{temp}
\end{equation}
By first law of thermodynamics, we can get  Hawking-Bekenstein entropy
\begin{equation}
\begin{split}
S&=\int_0^{r_+}\frac{1}{T}\left(\frac{\partial M}{\partial r_+}\right)dr_+=\pi r_+^2\\
&=\frac{A}{4}\label{entropy}
\end{split}
\end{equation}
$A$ stands for the horizon area. By employing Eq.(\ref{4}), black hole mass $M$ can be reexpressed as
\begin{equation}
M=\frac{r_+}{2}\left(1+\frac{Q^2}{r_+^2}-\frac{a}{r_+^{3\omega_q+1}}+\frac{8\pi Pr_+^2}{3}\right)\label{5}
\end{equation}
so the thermodynamic volume $V$ conjugated to the pressure $P$ can be derived from  Eq. (\ref{5})
\begin{equation}
V=\left(\frac{\partial M}{\partial P}\right)_{S,Q}=\frac{4\pi r_+^3}{3}
\end{equation}
To obtain Smarr relation, when quintessence field is taken into account the first law for RN-AdS black hole  must be modified as
\begin{equation}
dM=TdS+\phi dQ+VdP+\mathcal{A}da
\end{equation}
quantity $\mathcal A$
\begin{equation}
\mathcal{A}=\left(\frac{\partial M}{\partial a}\right)_{S,Q,P}=-\frac{1}{2r_+^{3\omega_{q}}}
\end{equation}
which is conjugated to $a$.  By dimension analysis, Smarr relation can be derived
\begin{equation}
M=2TS+\phi Q-2VP+(1+3\omega_{q})\mathcal{A}a
\end{equation}
We have following state equation
\begin{equation}
P=\frac{T}{\nu}-\frac{1}{2\pi \nu^2}+\frac{2Q^2}{\pi \nu^4}-\frac{8^{\omega_{q}}\times3a\omega_q}{\pi \nu^{3(1+\omega_q)}}\label{state}
\end{equation}
where $\nu=2r_+$ is the specific volume which actually is the double horizon radius of black hole, this differs with the specific volume with dimension [$L^3$] defined in traditional thermodynamics. What's more, we can obtain  the critical temperature $T_c$ and critical pressure $P_c$

\begin{gather}
T_c=\frac{1}{\pi \nu_c}-\frac{8Q^2}{\pi \nu_c^3}+\frac{8^{\omega_q}\times 9a\omega_q(1+\omega_q)}{\pi \nu_c^{3\omega_q+2}}\\
P_c=\frac{1}{2\pi \nu_c^2}-\frac{6Q^2}{\pi \nu_c^4}+\frac{8^{\omega_q}\times 3a\omega_q(2+3\omega_q)}{\pi \nu_c^{3(1+\omega_q)}}
\end{gather}
where $\nu_c$ is the critical specific volume which satisfies equation
\begin{equation}
\nu_c^2-24Q^2+\frac{9a\omega_q8^{\omega_q}(3\omega_q+2)(\omega_q+1)}{\nu_c^{3\omega_q-1}}=0\label{15}
\end{equation}
For $\omega_q=-\frac{2}{3}$, we have
\begin{equation}
\nu_c=2\sqrt{6}Q,\quad T_c=\frac{\sqrt{6}}{18\pi Q}-\frac{a}{2\pi}, \quad P_c=\frac{1}{96\pi Q^2}
\end{equation}
and when $\omega_q=-1$, we obtain
\begin{equation}
\nu_c=2\sqrt{6}Q,\quad T_c=\frac{\sqrt{6}}{18\pi Q}, \quad P_c=\frac{1}{96\pi Q^2}+\frac{3a}{8\pi}\label{9}
\end{equation}

For other values of state parameter $\omega_q$, the Eq. (\ref{15}) cannot be analytically solved so that we can not get the analytical formula for $\nu_c$, $T_c$ and $P_c$ and the numerical method must be adopted for help. On the other hand, we have ratio
\begin{equation}
\frac{P_c \nu_c}{T_c}\Big |_{\omega_q=-\frac{2}{3}}=\frac{3}{8-12\sqrt{6}aQ},\quad
\frac{P_c \nu_c}{T_c}\Big |_{\omega_q=-1}=\frac{3}{8}(1+36aQ^2)\label{16}
\end{equation}
which do not keep constant as the charged AdS-RN black hole in the spacetime without quintessence does but are dependent of factor $a$ and charge $Q$, and this result reflects the effects of the quintessence field on the black hole thermodynamics. One should note that when $a=0$, the ratio in Eq. (\ref{16}) will be reduced to
\begin{equation}
\frac{P_c \nu_c}{T_c}=\frac{3}{8}
\end{equation}
which keeps the same value as the van der Waals liquid-gas system. From another perspective, we can introduce a new temperature denoted as $w$ for $\omega_q=-\frac{2}{3}$
\begin{equation}
w=T_c+\frac{a}{2\pi}
\end{equation}
which is called the shifted temperature and the universal ratio value can be obtained as
\begin{equation}
\frac{P_c \nu_c}{w}=\frac{3}{8}
\end{equation}
and for $\omega_q=-1$, the similar procedure can be employed to obtain the universal ratio value.

\section{The efficiency of heat engines influenced by quintessence}
In this section, we will probe the effects of dark energy(i.e. quintessence field in this paper) on the efficiency of RN-AdS black holes as  heat engine defined in Fig \ref{fig3}. From Eq. (\ref{entropy}), (\ref{temp}) and (\ref{4}), Hawking temperature can be rewritten as
\begin{equation}
T=\frac{S^{-\frac{3}{2}(1+\omega_q)}\left(S^{\frac{3\omega_q}{2}}\left(-\pi Q^2+S+8PS^2\right)+3a\pi^{\frac{1+3\omega_q}{2}}\sqrt{S}\omega_q\right)}{4\sqrt{\pi}}\label{temp2}
\end{equation}
and then heat capacity can be calculated by employing Eq. (\ref{temp2})
\begin{equation}
C=T\frac{\partial S}{\partial T}=\frac{-2\left(-\sqrt{\pi}+2\frac{\partial P}{\partial T}\sqrt{S}\right)S\left(S^{3\omega_q/2}(-\pi Q^2+S+8PS^2)+3a\pi^{\frac{1+3\omega_q}{2}}\sqrt{S}\omega_q\right)}{\sqrt{\pi}S^{\frac{3\omega_q}{2}}\left(3\pi Q^2+S(8PS-1)\right)-3a\pi^{\frac{3\omega_q+2}{2}}\sqrt{S}\omega_q(3\omega_q+2)}\label{capacity}
\end{equation}
For heat capacity $C_P$ under the constant pressure $P$ can be obtained by condition
\begin{equation}
\frac{\partial P}{\partial T}=0\label{6}
\end{equation}
Substitute Eq. (\ref{6}) into Eq. (\ref{capacity}) to get $C_P$
\begin{equation}
C_P=\frac{2S^{\frac{3\omega_q+2}{2}}\left(S+8PS^2-\pi Q^2\right)+6a\pi^{\frac{3\omega_q+2}{2}}\sqrt{\frac{3}{2}}\omega_q}
{S^{\frac{3\omega_q}{2}}\left(3\pi Q^2+S(8PS-1)\right)-3a\pi^{\frac{3\omega_q+2}{2}}\sqrt{S}\omega_q(3\omega_q+2)}
\end{equation}
To get heat capacity $C_V$ under constant volume, we can substitute $\frac{\partial P}{\partial T}\big|_V$ into Eq. (\ref{capacity}) and we have
\begin{equation}
\frac{\partial P}{\partial T}\bigg|_V=\frac{\sqrt{\pi}}{2\sqrt{S}}\Rightarrow C_V=0
\end{equation}
The state equation \ref{state} and entropy formula \ref{entropy} are used in the calculations above. Both the RN-AdS black holes and the RN-AdS black holes surrounded by quintessence have the heat capacity $C_V=0$, which means that the quintessence would not affect the black hole heat capacity under the fixed thermodynamic volume while $C_P$ will be affected by quintessence field. Zero $C_V$ suggests that no heat exchange in the  isochoric process(2$\rightarrow$3 and 4$\rightarrow$1 in Fig \ref{fig3}) such that we only need to calculate heat $Q_H$ along the process $1\rightarrow2$ which can be computed as
\begin{equation}
\begin{split}
Q_H=\int_{T_1}^{T_2}C_P(P_1,T)dT&=\int_{r_1}^{r_2}C_P(P_1,T)\frac{\partial T}{\partial r}dr=
\frac{1}{6}\left(\frac{3\pi Q^2 +S_2(3+8P_1 S_2)}{\sqrt{\pi S_2}}-3a\pi^{\frac{3\omega_q}{2}}S_2^{-\frac{3\omega_q}{2}} \right)\\
&-\frac{1}{6}\left(\frac{3\pi Q^2 +S_1(3+8P_1 S_1)}{\sqrt{\pi S_1}}-3a\pi^{\frac{3\omega_q}{2}}S_1^{-\frac{3\omega_q}{2}} \right)
\end{split}
\end{equation}
the output work $W$, i.e. the area of rectangle is
\begin{equation}
W=\frac{4}{3\sqrt \pi}\left(S_2^{\frac{3}{2}}-S_1^{\frac{3}{2}}\right)(P_1-P_4)
\end{equation}
and now  the efficiency can be derived as
\begin{equation}
\eta=\frac{W}{Q_H}=\left(1-\frac{P_4}{P_1}\right)\times {\frac{1}{1+\frac{3(\sqrt{S_1S_2}-\pi Q^2)}{8\sqrt{S_1S_2}(S_1+S_2+\sqrt{S_1S_2})P_1}+\frac{3a\pi^{\frac{1+3\omega_q}{2}}(S_2^{-\frac{3\omega_q}{2}}-S_1^{-\frac{3\omega_q}{2}})}{8(S_1^{\frac{3}{2}}-S_2^{\frac{3}{2}})P_1}}}\label{effi}
\end{equation}
One can directly check that this formula indeed equals efficiency formula expressed by black mass $M$  in Eq. (\ref{14}).
Focusing on the large volume branch of solutions and therefore neglecting charge $Q$ to leading order, we can obtain
\begin{equation}
\eta=\left(1-\frac{P_4}{P_1}\right) \left\{1+\frac{1}{P_1}\left(\frac{3a\pi^{\frac{1+3\omega_q}{2}}(S_1^{-\frac{3\omega_q}{2}}-S_2^{-\frac{3\omega_q}{2}})}{8(S_1^{\frac{3}{2}}-S_2^{\frac{3}{2}})P_1}\right)
-\frac{3}{8P_1}\left(\frac{S_2^{\frac{1}{2}}-S_1^{\frac{1}{2}}}
{S_2^{\frac{3}{2}}-S_1^{\frac{3}{2}}}\right)+\mathcal{O}\left(\frac{1}{P_1^2}\right)\right\}\label{7}
\end{equation}
Considering the condition $a=0$ which means that the quintessence field disappears such that the metric (\ref{metric}) can be reduced to pure RN-AdS black holes, and then the efficiency (\ref{7}) would be simplified into
\begin{equation}
\eta=\left(1-\frac{P_4}{P_1}\right)\left\{1-\frac{3}{8P_1}\left(\frac{S_2^{\frac{1}{2}}-S_1^{\frac{1}{2}}}
{S_2^{\frac{3}{2}}-S_1^{\frac{3}{2}}}\right)+O\left(\frac{1}{P_1^2}\right)\right\}\label{8}
\end{equation}
which coincides with the result shown in Ref.\cite{Johnson1}. The normalization factor $a$ related to the density $\rho_q$ of the quintessence field by relation
\begin{equation}
\rho_q=-\frac{a}{2}\frac{3\omega_q}{r^{3(\omega_q+1)}}\label{density}
\end{equation}
Eq. (\ref{density}) implies that for the negative state parameter $\omega_q$, the factor $a$ must be limited to be positive to keep the density positive. For the entropy, we have $S_2>S_1$ and the negative $\omega_q$ will keep
$S_2^{-3\omega_q/2}>S_1^{-3\omega_a/2}$ which makes the second term in the brace of Eq.(\ref{7}) be positive such that the efficiency of the heat engine of the RN-AdS black holes surrounded by quintessence is higher than efficiency (Eq.\ref{8}) of the pure RN-AdS black holes as heat engine. In other words, the quintessence field, as a kind of dark energy model, can promote the heat engine efficiency of the charged AdS black holes provided that $\omega_q<0$. Considering that the term including a negative $\omega_q$(parameter $a$ is positive) in black hole mass formula (\ref{5}) has a negative contribution to $\partial M/\partial V$,  we find that our result is consistent with the result concluded in Ref.\cite{Hennigar1} which presents some general considerations for holographic heat engines for both static and rotating black holes, and states that for the black holes with $C_V=0$, any quantity contributing negatively to $\partial M/\partial V$ will lead to a larger efficiency.
Furthermore, the efficiency of the Carnot engine is
\begin{equation}
\eta_c=1-\frac{T_C}{T_H}=1-\frac{T_4(P_4,S_1)}{T_2(P_1,S_2)}\label{effic}
\end{equation}
After all the quantities we need for the discussion have been obtained above, it is necessary and important to investigate how the efficiency changes under the different normalization factor $a$, state parameter $\omega_q$, black hole entropy $S_2$ related to volume and the higher pressure $P_1$.

\begin{figure}[thbp]
\centering
\includegraphics[height=2.6in,width=3.2in]{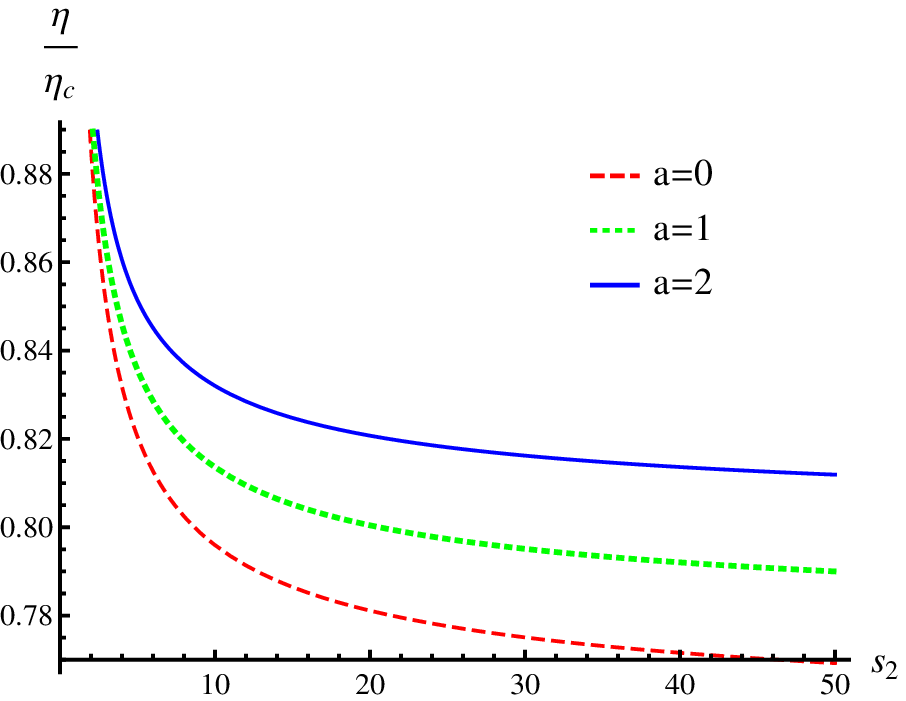}~~
\includegraphics[height=2.6in,width=3.2in]{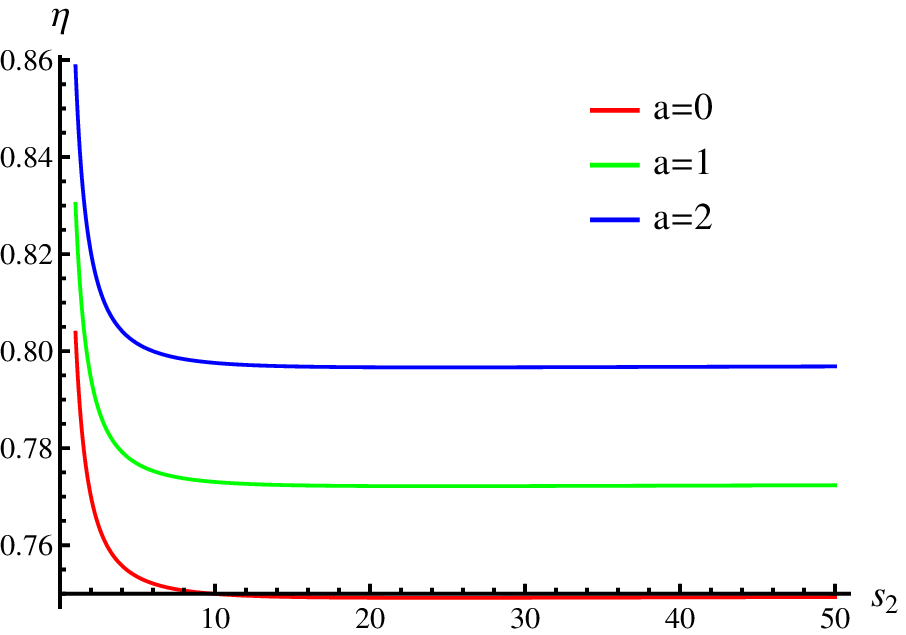}
\caption{The two  figures above are plotted at 3 different fixed normalization factor $a$ , here we take $P_1=4,P_4=1,S_1=1,Q=1$ and state parameter $\omega_q=-1$. The left figure corresponds to the ratio between efficiency $\eta$ and $\eta_c$, the right one to efficiency $\eta$\label{fig4}}
\end{figure}

We plot efficiency and $\eta$ and the ratio between $\eta$ and  Carnot efficiency $\eta_c$ by employing Eq. (\ref{effi}) and Eq. (\ref{effic}) in Fig \ref{fig4} which shows us that the heat engine efficiency  $\eta$ (right figure of Fig \ref{fig4}) monotonously decrease with the entropy $S_2$ (or volume $V_2$) increase implying that the bigger volume difference between the smaller black holes ($V_1$) and the bigger black holes($V_2$) corresponds to a lower efficiency. On the other hand, the decrease of the efficiency gradually become slower and slower as the volume difference becomes bigger till the efficiency to one constant number. What's more, we can see that when all the other parameters are fixed, the  heat engine with a bigger normalization factor $a$ always has the higher efficiency $\eta$, this fact tells us that the quintessence field with the bigger density $\rho_q$ would more significantly improve(or influence) the black hole heat engine efficiency. As we can see in the following parts of this present paper, the factor $a$  should be bounded in order to have a heat engine with the efficiency $\eta<1$. The left figure related to the $\frac{\eta}{\eta_c}$ of Fig \ref{fig4} behaves similar to the right one but their implications are distinct. We can see that $\frac{\eta}{\eta_c}<1$ always holds for 3 different $a$ at any  entropy $S_2$ and this result is consistent with the fact in the traditional thermodynamics that the Carnot heat engine has the highest efficiency otherwise the second law of the thermodynamics will be broken. On the other hand, the ratio $\eta/\eta_c$ monotonously decreases with the increase of entropy $S_2$ which suggests that the bigger the black hole is, the difference between the  actual efficiency $\eta$ and the idealized highest efficiency $\eta_c$ is bigger, in other words, the energy utilization rate is smaller.

\begin{figure}[thbp]
\centering
\includegraphics[height=2.6in,width=3.2in]{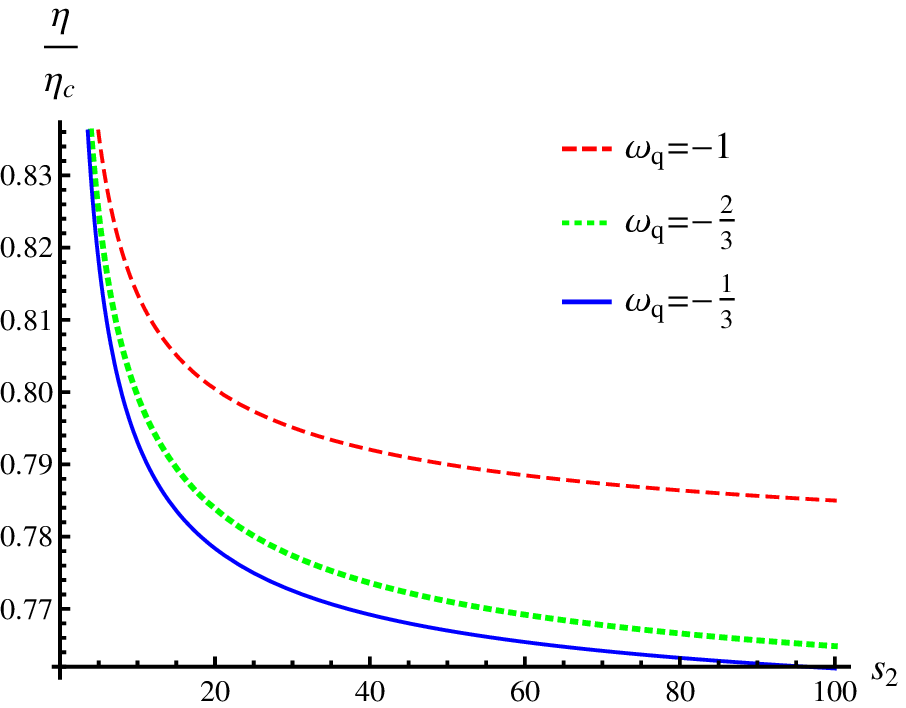}~~
\includegraphics[height=2.6in,width=3.2in]{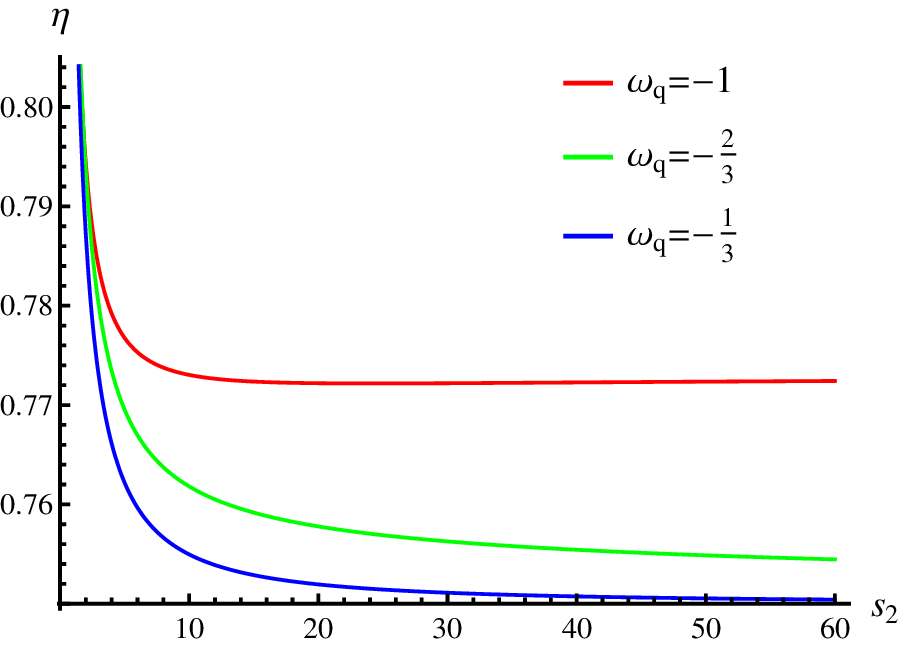}
\caption{The two  figures above are plotted at 3 different fixed state parameter $\omega_q$ , here we take $P_1=4,P_4=1,S_1=1,Q=1$ and normalization factor $a=1$. The left figure corresponds to the ratio between efficiency $\eta$ and $\eta_c$, the right one to efficiency $\eta$.\label{fig5}}
\end{figure}

Fig \ref{fig5} is plotted under three different state parameters $\omega_q$. The curves in this figure is highly similar to that in Fig \ref{fig4} therefor the similar discussions for Fig \ref{fig4} can be used in this situation but one point we should  point out. The Fig \ref{fig5} shows that for a bigger state parameter $\omega_q$ , the corresponding  heat engine efficiency is smaller. This result is apparent since the quintessence field density $\rho_q$ decrease with the parameter $\omega_q$ increases from $-1$ to $0$ at the constant factor $a$ as shown by Eq. (\ref{density}).

\begin{figure}[thbp]
\centering
\includegraphics[height=2.6in,width=3.2in]{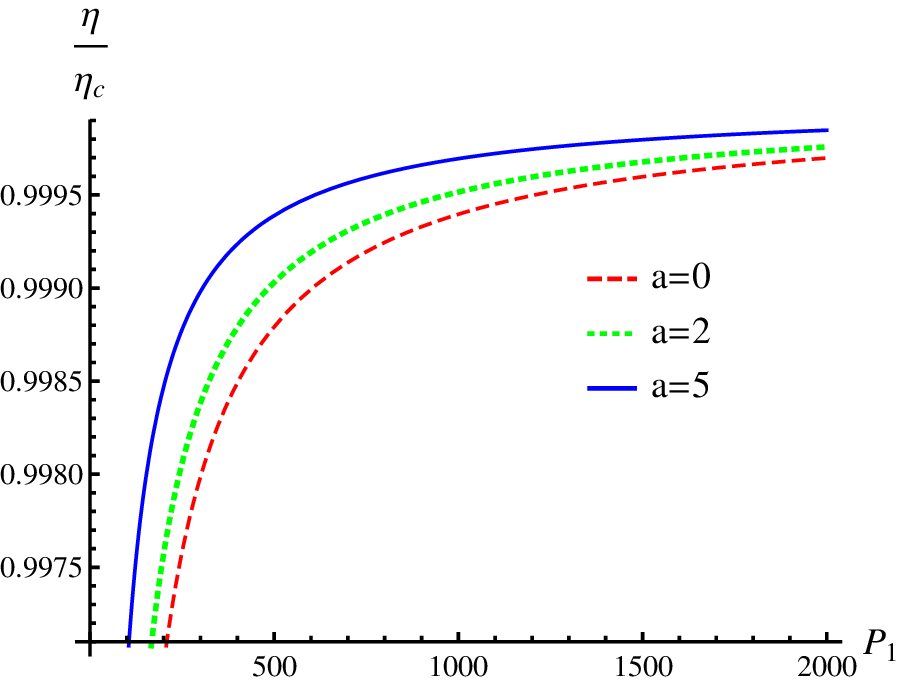}~~
\includegraphics[height=2.6in,width=3.2in]{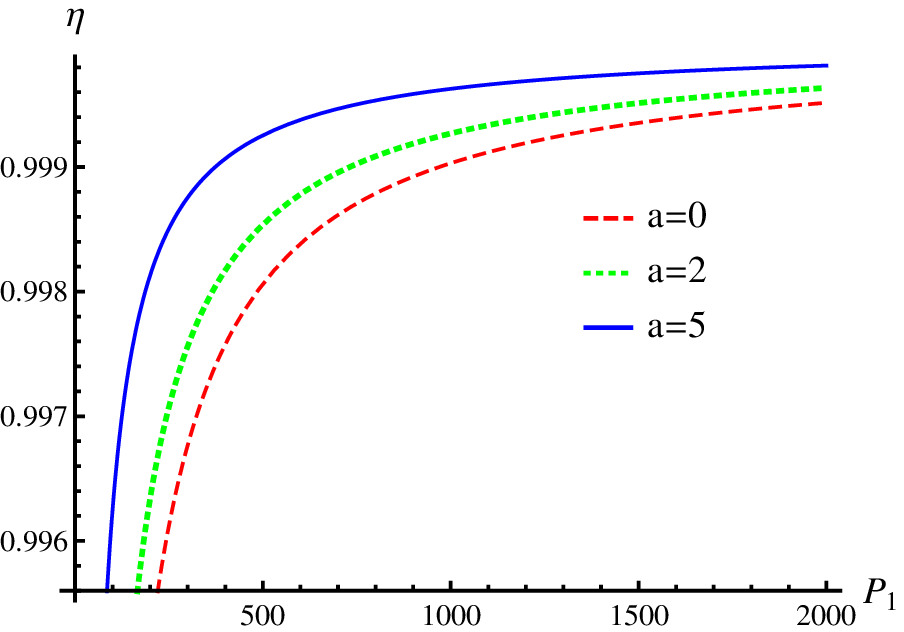}
\caption{The two  figures above are plotted at 3 different fixed normalization factor $a$ , here we take $P_2=4,S_1=1,S_2=4,Q=1$ and state parameter $\omega_q=-1$.\label{fig6}}
\end{figure}

On the other side, we  would like to investigate how the $\eta$ and $\eta_c$ change under the change of pressure $P_1$ by plotting Fig \ref{fig6}. We can see from the right figure that the efficiency monotonously increase as the $P_1$ increases but the efficiency would not approach  value 1 within the limited $P_1$.  The left figure shows that the efficiency would get closer and closer to the maximum efficiency value of the Carnot heat engine with the increase of $P_1$. Apparently, both $\eta$ and $\eta_c$ would infinitely approach 1 when $P_1$ goes to infinity.

\begin{figure}[thbp]
\centering
\includegraphics[height=2.6in,width=3.2in]{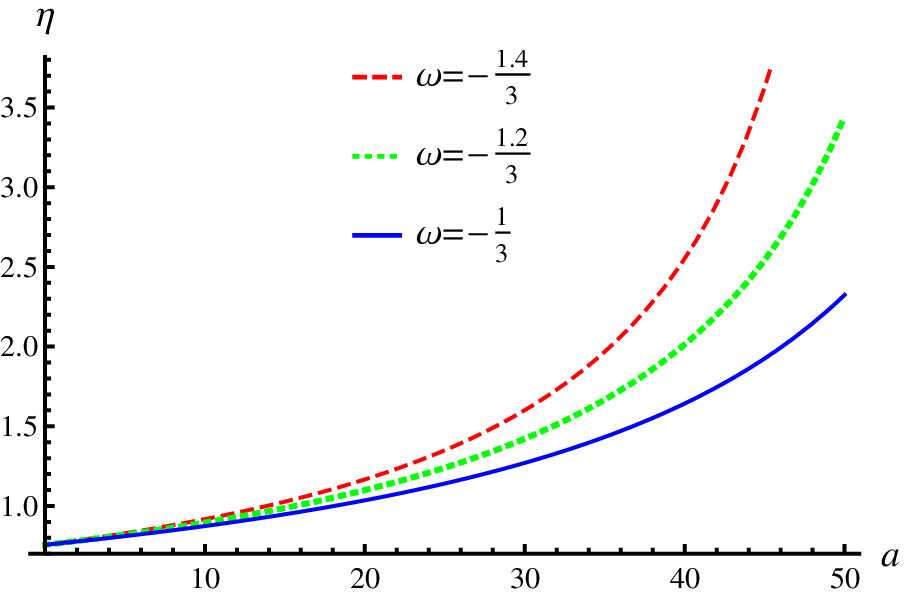}~~
\includegraphics[height=2.6in,width=3.2in]{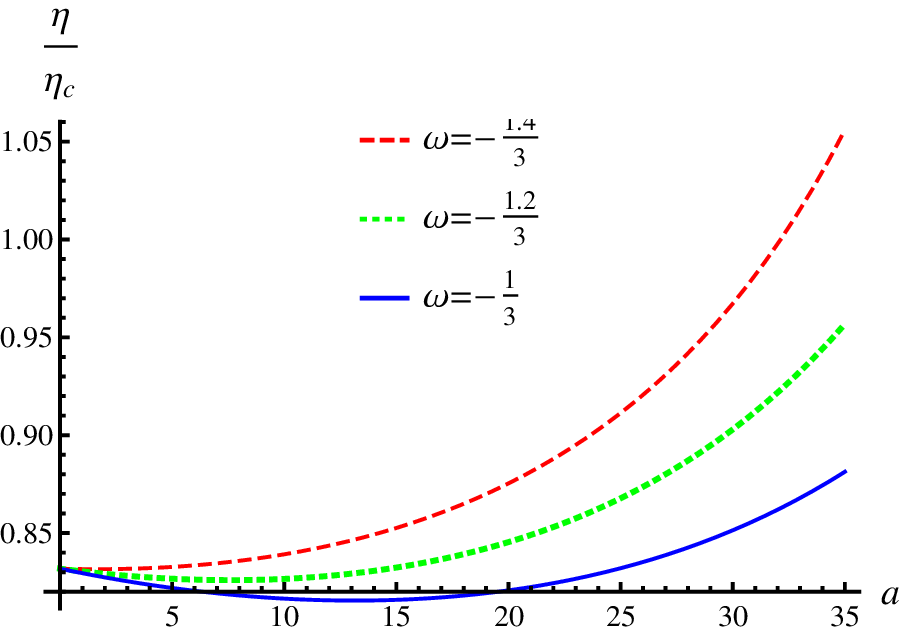}
\caption{The two  figures above are plotted at 3 different fixed state parameter $\omega_q$ , here we take $P_1=4,P_4=1,S_1=1,S_2=4,Q=1$.\label{fig7}}
\end{figure}

\begin{figure}[thbp]
\centering
\includegraphics[height=2.4in,width=3.2in]{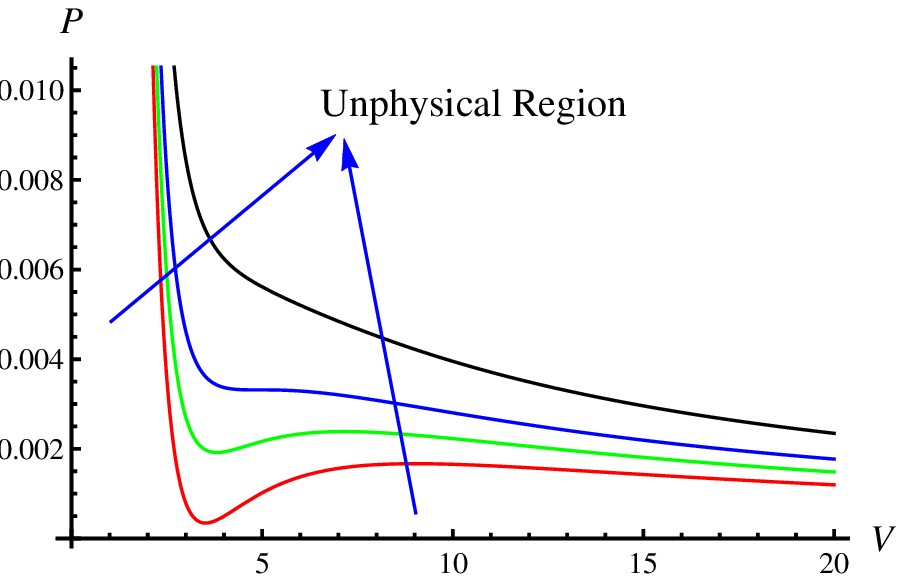}~~
\includegraphics[height=2.4in,width=3.2in]{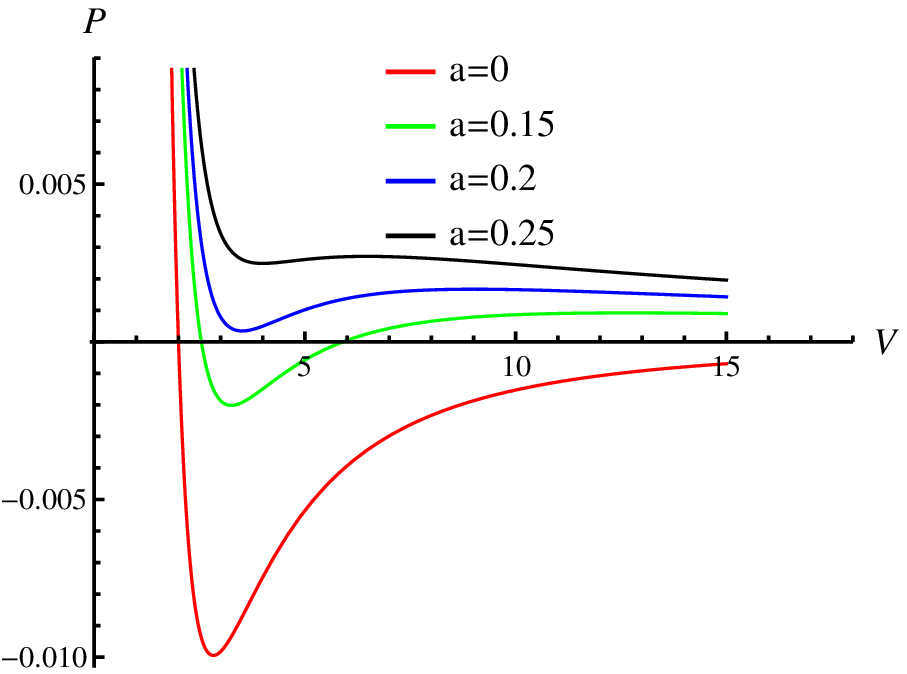}
\caption{For the left plot, the left region and the bottom area of the red isotherm which represents $T=0$ corresponds to  negative temperature isotherms which are unphysical. In the right plot, we plot 4 isotherms with temperature $T=0$, and we can see that the isotherms will move to the right as the increase of parameter $a$, i.e. the unphysical region will be shifted from left to the right when $a$ grows such that the cycle fixed in the plane will always be shifted to the unphysical region if $a$ is big enough, and this may serve as the origin for the $\eta>1$ and $\eta/\eta_c>1$ shown in Fig 7. Here we take $\omega_q=-\frac{2}{3}$. \label{fig12}}
\end{figure}

\begin{figure}[thbp]
\centering
\includegraphics[height=2.5in,width=3.8in]{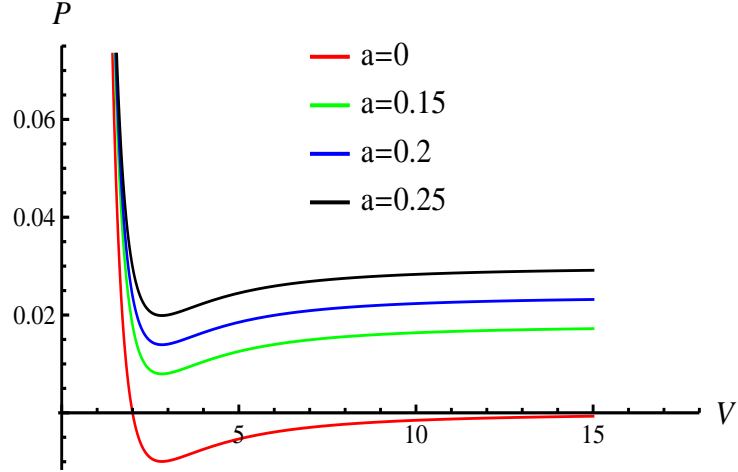}
\caption{The four $T=0$ isotherms are plotted in the case $\omega_q=-1$. With the increase of parameter $a$, the isotherms with fixed temperature($T=0$ here) will move to the upper region of the P-V plane such that the fixed thermodynamical cycle would  be shifted from the physical region to the unphysical region as the cycle with $\omega_q=-2/3$ does.\label{fig13}}
\end{figure}

Now it is time to probe how the quintessence field influence the efficiency. We plot figures of $\eta$ and $\eta/\eta_c$ in Fig \ref{fig7} under the change of the normalization factor $a$ at 3 fixed different state parameters. The figures show us that the efficiency $\eta$ monotonously increase with the increase of $a$ for all 3 $\omega_q$, while for $\eta/\eta_c$, only the curve corresponding to $\omega_q=-\frac{1.4}{3}$  monotonously increase
and the other two begin with decrease and then increase. One should note that both $\eta$ and $\eta/\eta_c$ would exceed the maximum value 1 and this phenomenon is forbidden to appear by laws of thermodynamics. In order to preserve the thermodynamics laws, the factor $a$ , or quintessence field density $\rho_q$ must be constrained in a range which can be roughly observed from Fig \ref{fig7}. In fact, the Eq. (\ref{temp}) shows that a big enough factor $a$ or electric charge $Q$ can lead to a negative temperature which is unphysical. On the other hand, since the thermodynamical cycle in Fig \ref{fig7} is fixed in the $P-V$ plane, the increase of the parameter $a$ will make the unphysical region in the $P-V$ plane move to the right so that the cycle will be shifted from the physical region to unphysical region and this plays a role in the occurrence of $\eta>1$ and $\eta/\eta_c>1$ as we show in Fig \ref{fig7}. The Fig \ref{fig12} and Fig \ref{fig13} are plotted to demonstrate the shift of the thermodynamical cycle fixed in the $P-V$ plane from physical region to unphysical region which serves as a possible reason discussed here for the unphysical value of $\eta$ and $\eta/\eta_c$.

\begin{figure}[thbp]
\centering
\includegraphics[height=2.6in,width=3.2in]{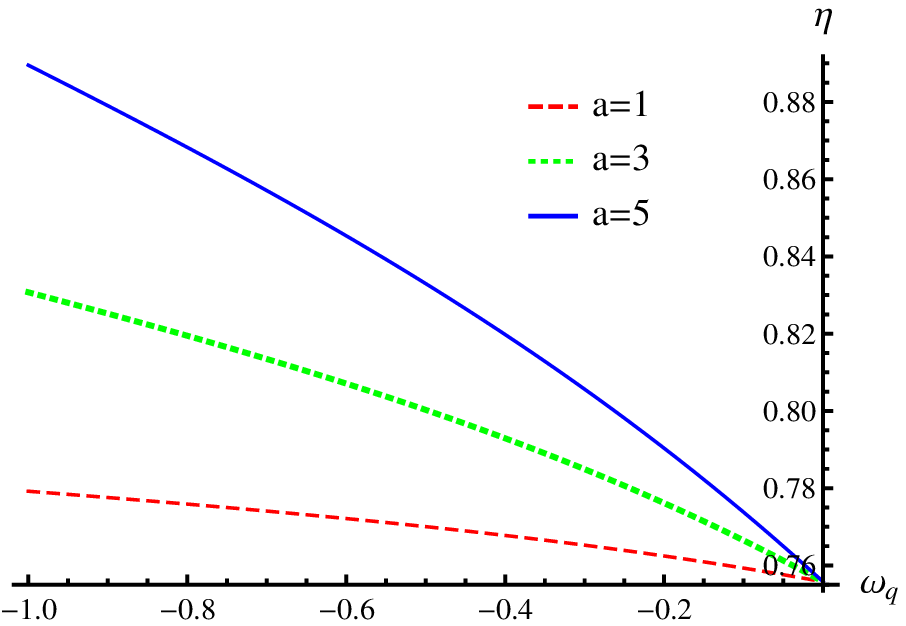}~~
\includegraphics[height=2.6in,width=3.2in]{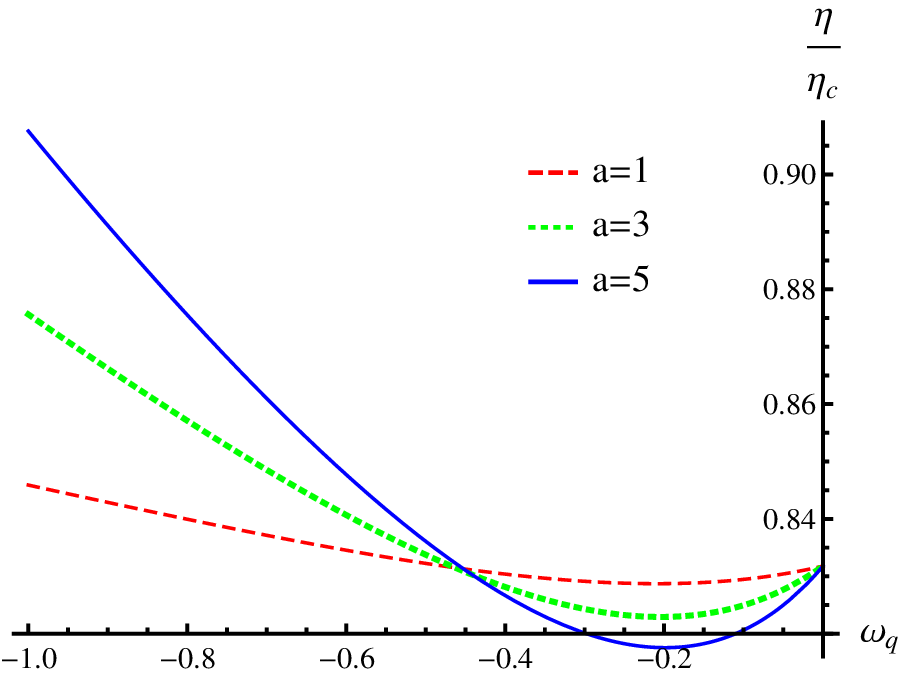}
\caption{The two  figures above are plotted at 3 different fixed normalization factor $a$ , here we take $P_1=4,P_4=1,S_1=1,S_2=4,Q=1$.\label{fig8}}
\end{figure}

Fig \ref{fig8} presents the efficiency $\eta$ and $\eta/\eta_c$ under the change of state parameter $\omega_q$.
As we discussed at the previous part of this paper, the higher field density $\rho_q$ corresponds to the higher efficiency as shown in the left figure of Fig \ref{fig8}. While for the value of $\eta/\eta_c$, which is bigger when the related factor $ a$  is bigger at the range $-1\leq\omega_q\leq-0.462945$ as portrayed by the right figure of Fig \ref{fig8}. But what interesting is that  at the range of $-0.462945\leq\omega_q\leq0$, we have  contrary result that
the bigger $a$ will lead to a smaller value of  $\eta/\eta_c$.

\begin{figure}[thbp]
\centering
\includegraphics[height=2.6in,width=3.2in]{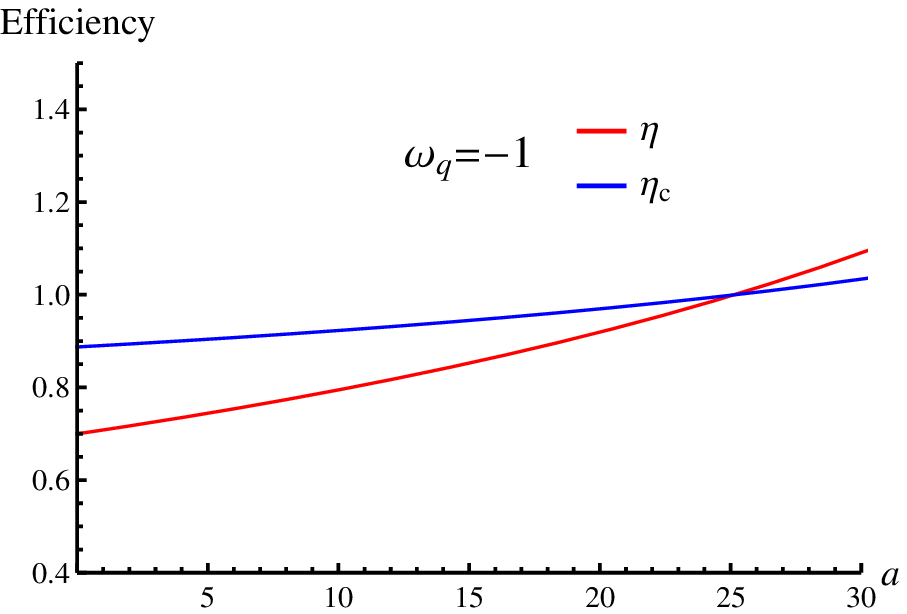}~~
\includegraphics[height=2.6in,width=3.2in]{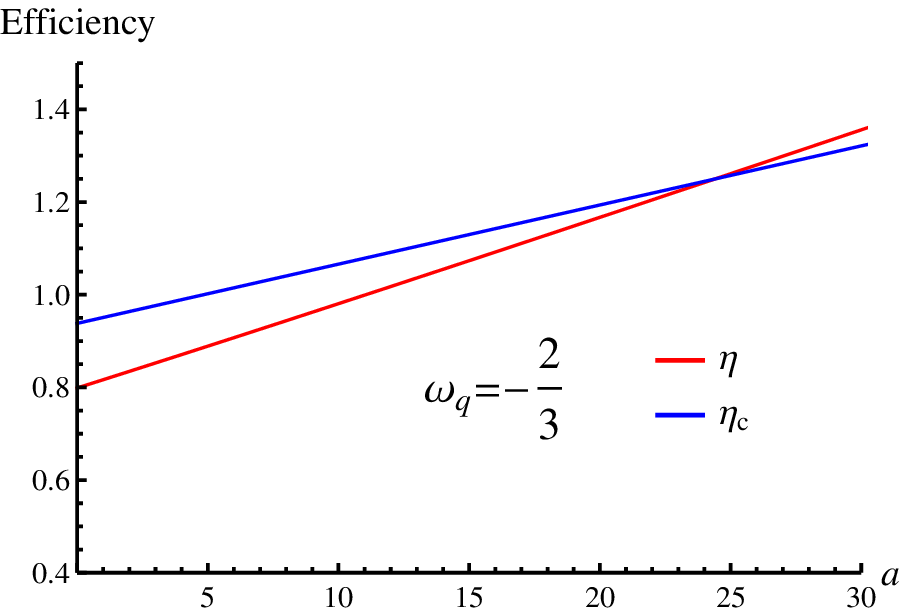}
\caption{a)The left plot is plotted by specifying $T_1=30$, $T_2=80$, $P_1=P_2=10$, $P_3=P_4=3$, charge $Q=1$ and state parameter $\omega_q=-1$. b)The right plot is plotted by  specifying $T_1=3$, $T_2=10$, $P_1=P_2=0.5$, $P_3=P_4=0.1$, charge $Q=1$ and state parameter $\omega_q=-\frac{2}{3}$. \label{fig17}}
\end{figure}

At this point, we have arrived at a conclusion that the quintessence field(as a famous scalar field model of dark energy) with negative state parameter $\omega_q$ can promote the efficiency of the heat engine defined in the extended phase space of the charged AdS black holes, but this result should be checked  further since  different schemes can be chosen and whether the quintessence field can improve efficiency of the heat engine operating in a new scheme is unknown yet. One choice of the new scheme is to specify the temperatures $T_1\& T_2$  and pressures $P_1\& P_4$ which can be used to calculate black hole mass hence the  Eq.(\ref{14}) is convenient to be  used to compute the efficiency. For simplicity, we only discuss the quintessence field with $\omega_q=-1$ and $\omega_q=-2/3$ due to that we can analytically solve the equation of state(Eq. \ref{state}) to get the value of mass. Fig \ref{fig17} is plotted to illustrate the behavior of the efficiency $\eta$ and $\eta_c$ which will increase with the grow of the normalization factor $a$(or dark energy density), and this result is consistent with the conclusion we have obtained in previous part of the present paper. From the plots we see that factor $a$ should be restricted to keep the efficiency physical, i.e. $\eta<\eta_c$.

\section{Critical black holes as heat engines}

\begin{figure}[thbp]
\centering
\includegraphics[height=2.6in,width=3.5in]{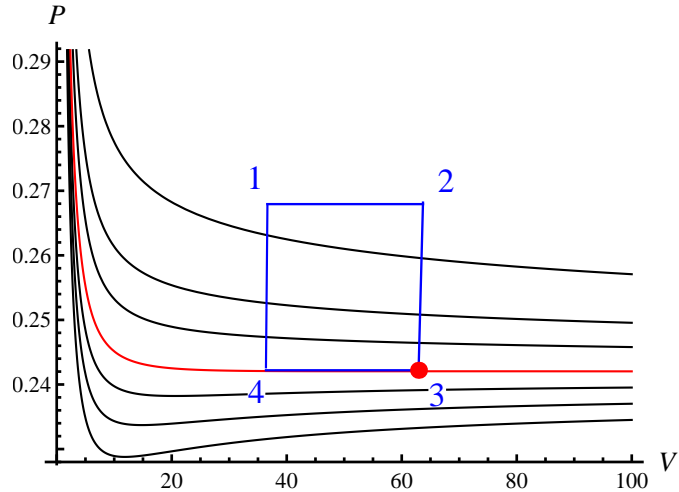}
\caption{The corner 3 of the  engine cycle is placed at the critical point(i.e. $T_c=0.0433$, $P_c=0.242$ and $V_c=61.56$), and the red curve stands for the $T=T_c$ isotherm, here we take $a=2,Q=1$.\label{fig16}}
\end{figure}

The latest progress of the black hole heat engine research made by Johnson \cite{Johnson5} demonstrates that the black hole heat engine defined in the $P-V$ plane  with the thermodynamical quantities at critical point  can approach the Carnot limit with the increase of electric charge $Q$ at finite power is possible. This discovery is amazing since the Carnot heat engine  setting an upper bound on the efficiency of a heat engine is an ideal, reversible engine of which a single cycle must be performed in infinite time which is impractical and so the Carnot engine has zero power. Inspired by this work, we find that the RN-AdS black hole surrounded by quintessence with the state parameter $\omega_q=-1$ can be treated in similar way in which one can solve the problem shown in Fig \ref{fig7}, i.e. the $\eta>1$ and $\eta/\eta_c>1$. The core of this way is to find a thermodynamic cycle which keeps the temperature non-negative at any given $a$ or electric charge $Q$. To achieve this goal, the position of the cycle should be changed with the change of the parameter $a$ or charge $Q$ rather than be fixed in the $P-V$ plane, and thereby the scaling behavior of the thermodynamical quantities must be used to define the thermodynamical cycle. We here would use the critical pressure $P_c$, critical temperature $T_c$ and critical specific volume $\nu_c$ with the state parameter  $\omega_q=-1$ in Eq. (\ref{9}). There are several ways of making such a choice which is essential and crucial to the property of the heat engines as we can see in the following discussions, and just two families are chosen here for illustration. Fig \ref{fig16} is the heat engine constructed near the critical point we are going to investigate in this section.

\subsection{Heat engine model 1}
In this section, we  construct the heat engines with the following choice
\begin{align}
&V_2=V_3=V_c=\frac{4\pi}{3}\left(\frac{\nu_c}{2}\right)^3,\quad P_3=P_4=P_c,\quad P_1=P_2=\frac{3}{2}P_c\label{10}\\
&V_1=V_4=\alpha V_c, \quad S_2=S_3=\pi \left(\frac{3V_c}{4\pi}\right)^{\frac{2}{3}},\quad S_1=S_4=\pi \left(\frac{3\alpha V_c}{4\pi}\right)^{\frac{2}{3}}\label{11}
\end{align}
where the dimensionless factor $\alpha$ is a constant restricted to $0<\alpha<1$ to keep $V_1=V_4$ smaller than $V_c$. By employing Eq.(\ref{10}), (\ref{11}), (\ref{9}) and (\ref{7}), the efficiency $\eta$ can be reduced to
\begin{equation}
\eta=\frac{(1+36aQ^2)(\alpha^{1/3}+\alpha^{2/3}+\alpha)}{3(12aQ^2+5)\alpha^{1/3}+3(1+12aQ^2)(\alpha+\alpha^{2/3})-2}\label{17}
\end{equation}
and the Carnot efficiency is
\begin{equation}
\eta_c=1-\frac{T_4(S_4,P_4)}{T_2(S_2,P_2)}=\frac{2-12\alpha^{2/3}+19\alpha+108aQ^2\alpha-6\alpha^{4/3}}{19\alpha+108aQ^2\alpha}\label{18}
\end{equation}

\begin{figure}[thbp]
\centering
\includegraphics[height=2.6in,width=3.2in]{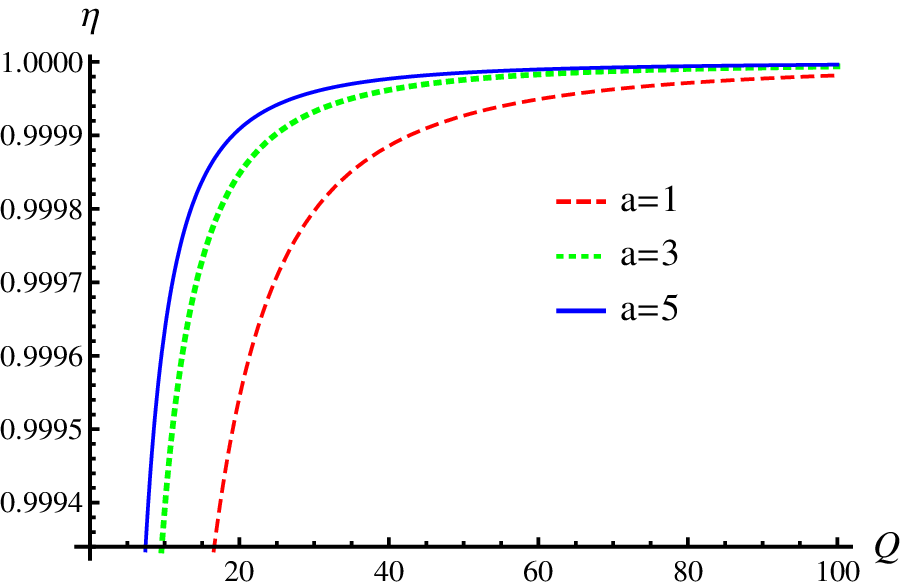}~~
\includegraphics[height=2.6in,width=3.2in]{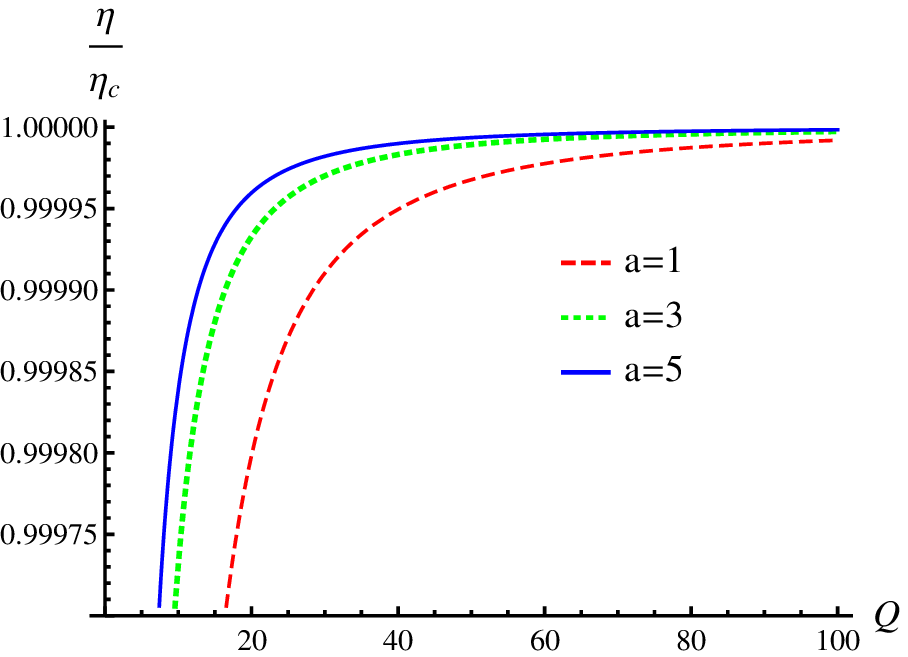}
\caption{The two  figures above are plotted under the change of the electric charge $Q$ at 3 different fixed normalization factor $a$, here we take $\alpha=\frac{1}{8}$.\label{fig9}}
\end{figure}

\begin{figure}[thbp]
\centering
\includegraphics[height=2.6in,width=3.2in]{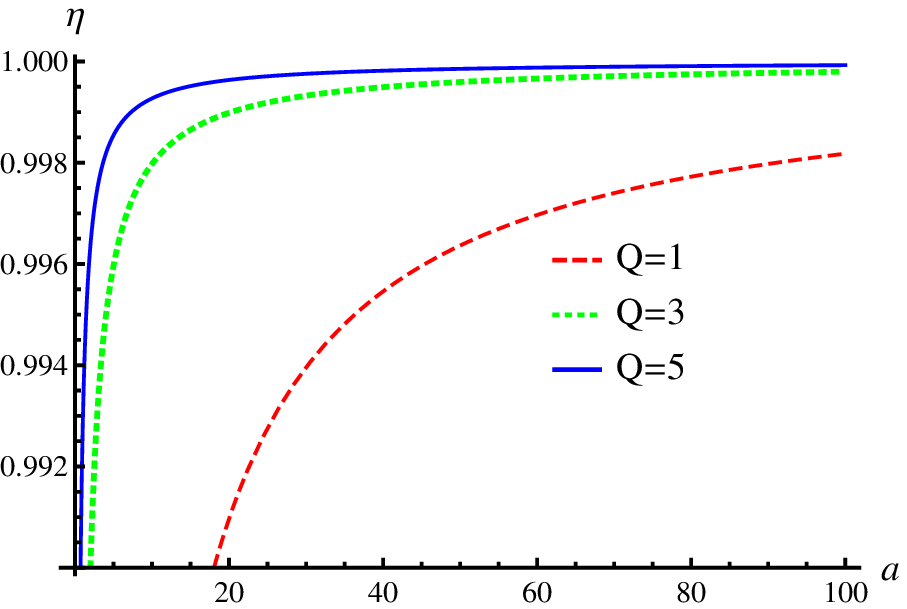}~~
\includegraphics[height=2.6in,width=3.2in]{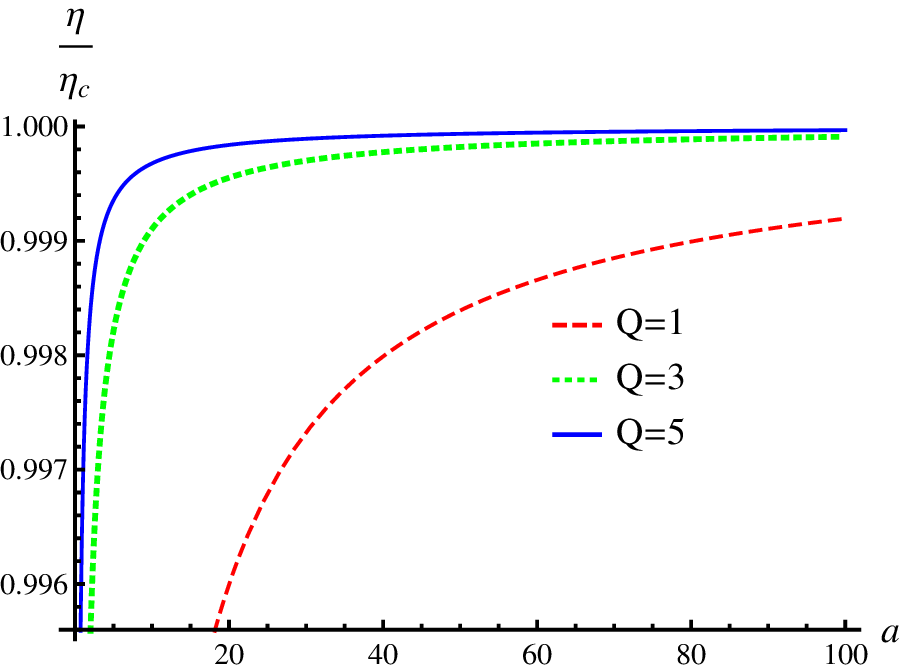}
\caption{The two  figures above are plotted under the change of the normalization factor $a$ at 3 different fixed electric charge $Q$, here we take $\alpha=\frac{1}{8}$.\label{fig10}}
\end{figure}

To have a better understanding of how the charge $Q$ influences efficiency, we  carry out large $Q$ expansion for both $\eta$ and $\eta_c$
\begin{gather}
\eta=1-\frac{23}{126aQ^2}+\frac{1219}{31752a^2Q^4}-\frac{64607}{8001504a^3Q^6}+\mathcal{O}(Q^{-7})\\
\eta_c=1-\frac{11}{108aQ^2}+\frac{209}{11664a^2Q^4}-\frac{3971}{1259712a^3Q^6}+\mathcal{O}(Q^{-7})
\end{gather}
where we have taken $\alpha=\frac{1}{8}$ for simplicity. The large $Q$ expansions for $\eta$ and $\eta_c$ indicate that  both $\eta$ and $\eta_c$ will infinitely approach value 1 when $Q$ goes to infinity, and no unphysical efficiency values appear.
We plot $\eta$ and $\eta/\eta_c$ in Fig \ref{fig9} under the change of electric Q. The figures show that the efficiency $\eta$ and $\eta/\eta_c$ increase with the grow of the electric charge, and $\eta$ will  approach the Carnot efficiency $\eta_c$ which is the maximum value of the efficiency and suggests that approach the Carnot limit is possible as what Johnson found in Ref.\cite{Johnson5}.
In practice, on account of the electric charge $Q$ should be finite, so that the efficiency $\eta<1$   which is allowed by thermodynamics laws, and increasing $Q$ is supposed to be a way to promote the heat engine efficiency.
Furthermore, we would like to expand $\eta$ and $\eta_c$ at large parameter $a$ with $\alpha=\frac{1}{8}$
\begin{gather}
\eta=1-\frac{23}{126aQ^2}+\frac{1219}{31752a^2Q^4}-\frac{64607}{8001504a^3Q^6}+\mathcal{O}({a^{-4}})\\
\eta_c=1-\frac{11}{108aQ^2}+\frac{209}{11664a^2Q^4}-\frac{3971}{1259712a^3Q^6}+\mathcal{O}(a^{-4})
\end{gather}
Interestingly, the expansion formulas at large $a$ are exactly the same as the formulas of the large $Q$ expansion, and this coincidence arises from the fact that parameter $a$ and charge $Q$ are always bonded together  in the form of $aQ^2$.
We draw the diagram of $\eta$ and $\eta/\eta_c$ under the change of factor $a$, as we plot in Fig \ref{fig10} and no unphysical efficiency values appear. So far, the problem of $\eta>1$ and $\eta/\eta_c>1$ demonstrated by Fig \ref{fig7} for large parameter $a$  has been solved by making some special choices of thermodynamical quantities listed in Eq. (\ref{10}) and (\ref{11}) near critical point.

\begin{figure}[thbp]
\centering
\includegraphics[height=2.6in,width=3.2in]{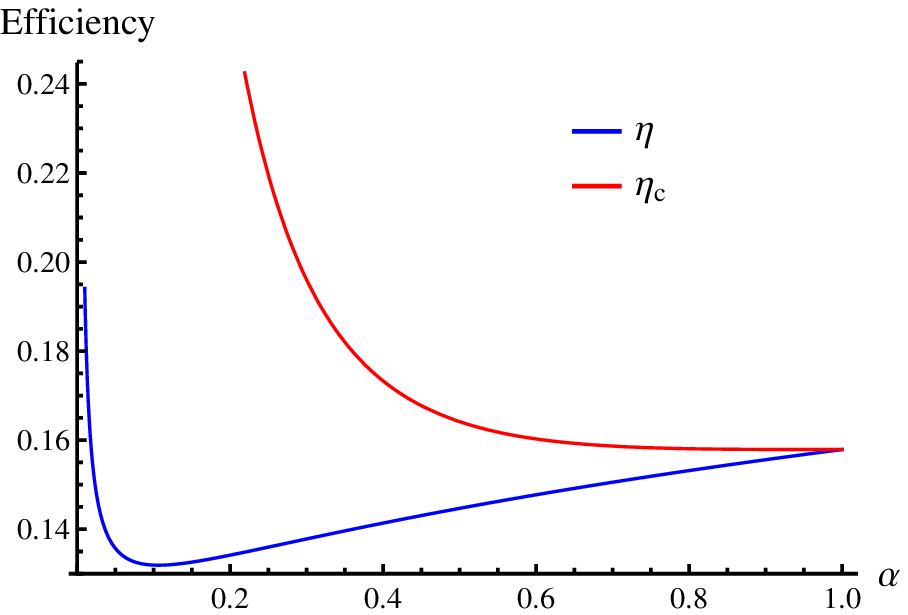}~~
\includegraphics[height=2.6in,width=3.2in]{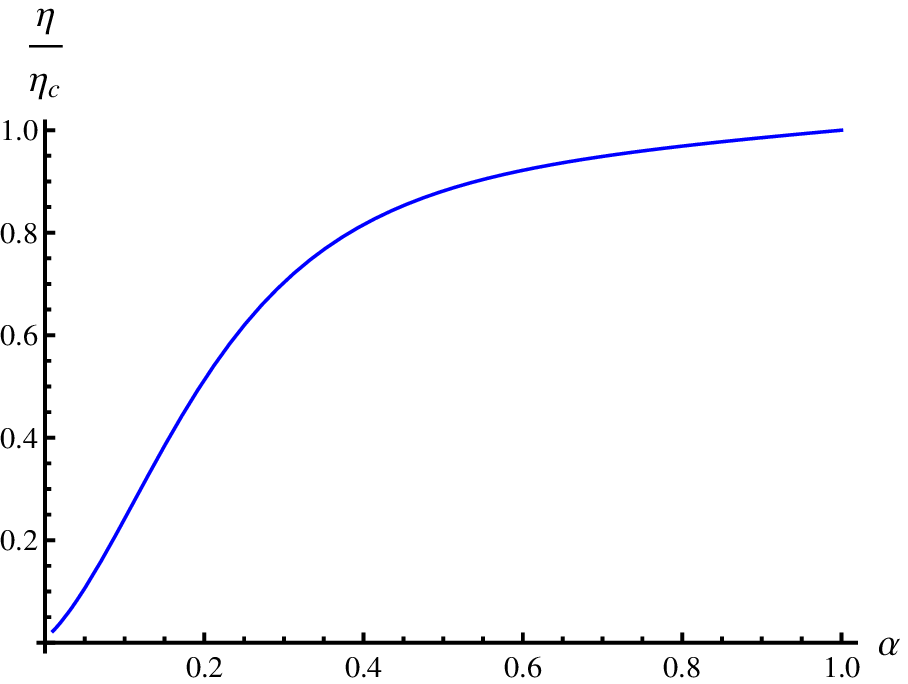}
\caption{The behavior of efficiency $\eta$, Carnot efficiency $\eta_c$ and the ratio $\frac{\eta}{\eta_c}$ under the condition of $a=0$.\label{fig14} }
\end{figure}

To demonstrate how the approach to Carnot efficiency is affected by the presence of the quintessence field,it would be useful here to discuss the heat engines when normalization factor $a=0$(i.e. the quintessence field is absent) under the choice of the thermodynamical quantities listed in Eq. (\ref{10}) and (\ref{11}) which generate a heat engine with the efficiency which is independent of charge $Q$. From Eq.(\ref{17}) and (\ref{18}), it is directly to get  $\eta$ and $\eta_c$ in the condition of $a=0$
\begin{gather}
\eta=\frac{\alpha^{1/3}+\alpha^{2/3}+\alpha}{3(5\alpha^{1/3}+\alpha^{2/3}+\alpha)-2}\\
\eta_c=\frac{2-12\alpha^{2/3}+19\alpha-6\alpha^{4/3}}{19\alpha}
\end{gather}
Obviously, both $\eta$ and $\eta_c$  are  constant and only depend on the dimensionless constant factor $\alpha=\frac{V_1}{V_2}$. One should be aware  that factor $\alpha$ must be chosen appropriately to avoid the  occurrence of unphysical value of efficiency. Once $\alpha$ is selected, the efficiency is a constant which can not be influenced by the charge $Q$ if the quintessence field is absent(i.e. $a=0$), otherwise the efficiency will be dependent of normalization factor $a$ and charge $Q$ which provide a possible way to approach the Carnot limit at finite power. The Fig \ref{fig14} demonstrates the behavior of efficiency and the ratio $\eta/\eta_c$ under the change of $\alpha$. There is a problem should be pointed out in the left plot of Fig \ref{fig14} as one can find that when $\alpha=1$, the efficiency $\eta$ and Carnot efficiency $\eta_c$ has the relation
\begin{equation}
\eta=\eta_c=\frac{3}{19}
\end{equation}
However, the $\alpha=1$ means that $V_1=V_2$ which will lead to a zero work $W$ which can be calculated as
\begin{equation}
W=(P_1-P_4)(V_2-V_1)=\frac{Q(1-\alpha)}{4\sqrt{6}}=0,\quad (\alpha=1)
\end{equation}
while at the same time, we have
\begin{equation}
Q_H=-\frac{Q(2+\alpha^{1/3}(12\alpha^{1/3}+3\alpha-17))}{4\sqrt{6}\alpha^{1/3}}=0,\quad (\alpha=1)
\end{equation}
and the limit of $\alpha\rightarrow1$ will give us the nonzero and finite efficiency
\begin{equation}
\eta=\eta_c=\lim_{\alpha\rightarrow1}\frac{W}{Q_H}=\frac{3}{19}
\end{equation}
although the area of engine cycle has shrunk to zero. The problem  arises from the work $W$ done by heat engine is not fixed as a constant and the next  section is devoted to investigate a new heat engine with thermodynamical cycle which has constant area such that the problem occurs above could be avoided.

\subsection{Heat engine model 2}
\begin{figure}[thbp]
\centering
\includegraphics[height=2.6in,width=3.2in]{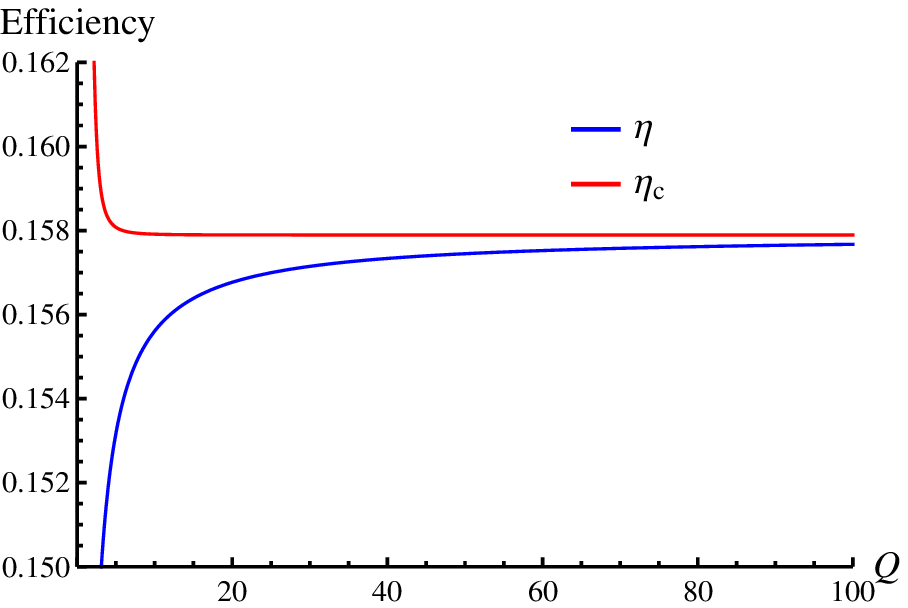}~~
\includegraphics[height=2.6in,width=3.2in]{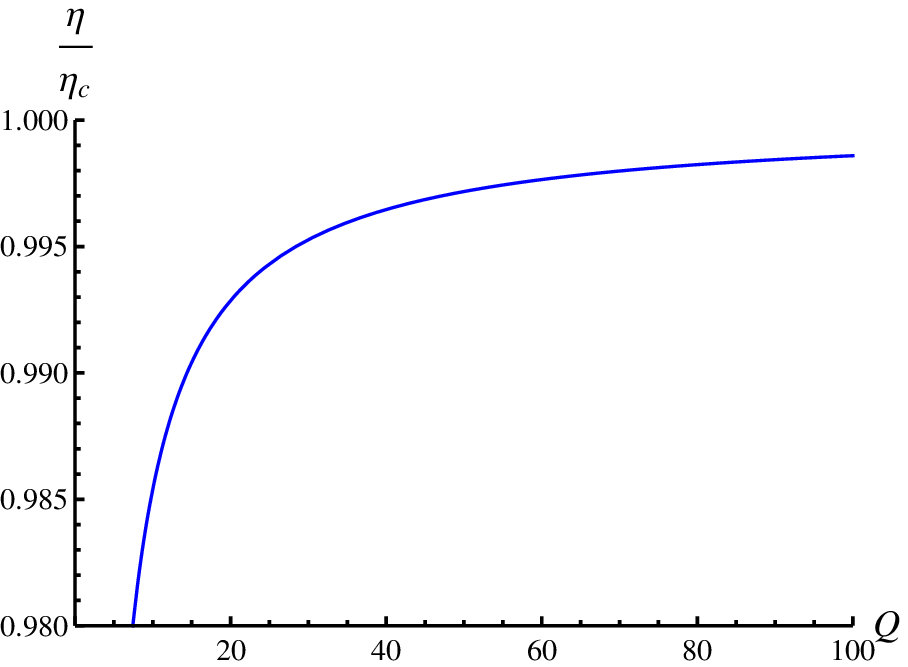}
\caption{The behavior of efficiency $\eta$, Carnot efficiency $\eta_c$ and the ratio $\frac{\eta}{\eta_c}$ as the function of $Q$.\label{fig15} }
\end{figure}

The heat engines discussed in this section is constructed by the following choices of thermodynamical quantities as
\begin{align}
&V_2=V_3=V_c=\frac{4\pi}{3}\left(\frac{\nu_c}{2}\right)^3,\quad P_3=P_4=P_c,\quad P_1=P_2=\frac{3(1+24aQ^2)}{2(1+36aQ^2)}P_c\label{19}\\
&V_1=V_4=V_c(1-\frac{L}{Q}), \quad S_2=S_3=\pi \left(\frac{3V_c}{4\pi}\right)^{\frac{2}{3}},\quad S_1=S_4=\pi \left(\frac{3V_1}{4\pi}\right)^{\frac{2}{3}}\label{20}
\end{align}
where $L$ is a constant and is restricted to $0<\frac{L}{Q}<1$, and the work is
\begin{equation}
W=(P_1-P_4)(V_2-V_1)=\frac{L}{4\sqrt{6}}
\end{equation}
which is also a constant. We have the efficiency
\begin{gather}
\eta=\frac{Q+Q^{2/3}(Q-L)^{1/3}+Q^{1/3}(Q-L)^{2/3}-L}{Q+15Q^{2/3}(Q-L)^{1/3}+3Q^{1/3}(Q-L)^{2/3}-3L}
\end{gather}
and the Carnot efficiency is
\begin{equation}
\eta_c=\frac{19L-6L(1-L/Q)^{1/3}-21Q+6Q^{2/3}(Q-L)^{1/3}+12Q^{1/3}(Q-L)^{2/3}}{19(L-Q)}
\end{equation}
This result  is interesting since both $\eta$ and $\eta_c$ are independent of normalization factor $a$ and only depend on the constant $L$ and charge $Q$ implying that the effects of the quintessence field on the efficiency of the heat engine can be eliminated by employing the thermodynamical quantities listed in Eq.(\ref{19}) and (\ref{20}). It is not hard to check that the efficiency $\eta$ and $\eta_c$ we calculate here are equivalent to the efficiency calculated for pure RN-AdS black holes as discussed in Ref.\cite{Johnson5}.
We have  expansion of efficiency at large $Q$
\begin{equation}
\eta=\frac{3}{19}-\frac{8L}{361Q}-\frac{416L^2}{61731Q^2}-\frac{3286L^3}{1172889Q^3}+\mathcal{O}(Q^{-4})
\end{equation}
and
\begin{equation}
\eta_c=\frac{3}{19}+\frac{8L^3}{513Q^3}+\frac{14L^4}{513Q^4}+\mathcal{O}(Q^{-5})
\end{equation}
These two expansions coincide with the results shown in Ref.\cite{Johnson5} and indicate that the efficiency $\eta$ will approach Carnot limit when charge $Q$ goes to infinity and the Fig \ref{fig15} is plotted to show the behavior of  $\eta$, $\eta_c$ and the ratio $\eta/\eta_c$ under the change of $Q$. The this model also reflects the fact that the choices of the thermodynamical quantities used to construct the heat engine is essential and crucial for the property of the heat engine as what we find in this section that the engine efficiency does not depend on the quintessence field which plays a role in influencing the efficiency for the engine discussed previously in this paper.

\section{Conclusions and Discussions }
In this paper, we study the heat engine where charged AdS black holes surrounded by dark energy is the working substance and the mechanical work is done via the $PdV$ term of the first law of black hole thermodynamics in the extended phase space. We first investigate the effects of a kind of dark energy (quintessence field in this paper) on the efficiency of the RN-AdS black holes as heat engine defined as a rectangle closed path in the $P-V$ plane. We find that quintessence field can improve the heat engine efficiency which will increase as the field density $\rho_q$ grows. Under some fixed parameters, we find that  bigger volume difference between the
smaller black holes($V_1$) and the bigger black holes($V_2$ ) will lead to a lower efficiency, and the bigger pressure difference will make the efficiency higher but  the efficiency is always smaller than 1 and will never be bigger than Carnot heat engine, this is consistent to the heat engine nature in traditional thermodynamics. With the pressure difference $P_1-P_4$ goes to infinity, the efficiency will approach value 1 and meanwhile it will also approach Carnot limit which is the maximum value of the heat engine efficiency allowed by thermodynamical laws. We find that for some large normalization factor $a$ (or large quintessence field density $\rho_q$ ) and electric charge $Q$ can make some thermodynamical quantities  unphysical and subsequently these unphysical quantities lead to the efficiency $\eta>1$ and $\eta>\eta_c$ which is not permitted by nature laws, and to avoid this problem both the factor $a$ and $Q$ may should be restricted to a certain range. We discuss the efficiency  under the change of the state parameter $\omega_q$. With the help of Fig \ref{fig8}, we find that with the increase of $\omega_q(-1\rightarrow0)$ the efficiency decrease monotonously. For $\eta/\eta_c$, it will decrease first and then increase with the grow of $\omega_q$, and the bigger $a$ corresponds to a bigger value of  $\eta/\eta_c$ at the beginning, while $\omega_q$ beyond one ceratin value(about -0.462945), the result is on the contrary. At last, inspired by Johnson's latest work \cite{Johnson5}, we find that the factor $a$ and electric charge $Q$ need not to be restricted if we put the thermodynamics cycle in a proper position(near the critical point) in $P-V$ plane by using the critical thermodynamical quantities calculated when state parameter  $\omega_q=-1$. With this method, we build two different heat engine model by making two different choices of thermodynamical quantities near the critical point. In model 1, the work done by the heat engine  determined by  E.q (\ref{10}) and  (\ref{11}) is not fixed. The efficiency $\eta$ and $\eta/\eta_c$ would increase with the grow of both charge $Q$ and $a$ and the heat engine will approach Carnot limit when $a$ or $Q$ goes to infinity. When $a=0$, we find that the efficiency is independent of charge $Q$ and only depend on the constant ratio $\alpha$ between $V_1$ and $V_2$. In model 2, we make the work a constant and find that
the efficiency is independent of quintessence field , in other words, the quintessence can not impact the efficiency of the heat engine provided that  which is constructed by choices of Eq.(\ref{19}) and (\ref{20}). The efficiency will also approach the Carnot limit when $Q$ goes to infinity as heat engine model 1 does. The different characteristics between the  two models also reflect the importance of the choice of the thermodynamical cycle to the property of the heat engines.

This paper have showed the impact of dark energy on the efficiency of 4 dimensional static charged AdS black holes as heat engine, and it would be of great interest to extend our present investigation to black holes in higher dimensions, and rotating black holes are also very worthwhile to investigate in the quintessence
AdS black hole spacetime. 
It is worth noting that the dual field theoretical interpretation about the quintessence is still unclear, although some attentions \cite{Zeng2,Zeng1,Chen3} have been paid to the holographic understanding in the framework of quintessence dark energy.
These problems and our results(especially we find that the quintessence dark energy can improve the efficiency of the holographic heat engines) are meaningful since they may help us promote our  understanding of black hole thermodynamics and the dual field theoretical interpretation of quintessence dark energy, further disclosing the relationship between dark energy and black holes. We leave these problems  for future work.

\section*{Acknowledgements}
This project is partially supported by NSFC. Hang Liu would like to express gratitude to Yan-Fei Tian for her great help in the process of getting this work finished.
\bibliography{HE}

\providecommand{\href}[2]{#2}\begingroup\raggedright\begin{thebibliography}{10}

\bibitem{Hawking1}
S.~Hawking and D.~N. Page, \emph{Thermodynamics of {B}lack {H}oles in anti-{D}e
  {S}itter {S}pace}, {\emph{Commun.Math.Phys.} {\bf 87} (1983) 577}.

\bibitem{Kubiznak1}
D.~Kubiznak and R.~B. Mann, \emph{${P}-{V}$ criticality of charged {A}d{S}
  black holes}, {\emph{JHEP} {\bf 1207} (2012) 033}.

\bibitem{Xu1}
J.~Xu, L.-M. Cao and Y.-P. Hu, \emph{${P}-{V}$ criticality in the extended
  phase space of black holes in massive gravity}, {\emph{Phys.Rev.D} {\bf 91}
  (2015) 124033}.

\bibitem{Cai1}
R.-G. Cai, L.-M. Cao, L.~Li and R.-Q. Yang, \emph{${P}-{V}$ criticality in the
  extended phase space of gauss-bonnet black holes in ads space}, {\emph{JHEP}
  {\bf 2013} (2013) 1--22}.

\bibitem{Banerjee1}
R.~Banerjee and D.~Roychowdhury, \emph{Critical behavior of born-infeld ads
  black holes in higher dimensions},
  \href{http://dx.doi.org/10.1103/PhysRevD.85.104043}{\emph{Phys. Rev. D} {\bf
  85} (May, 2012) 104043}.

\bibitem{Banerjee2}
R.~Banerjee and D.~Roychowdhury, \emph{Critical phenomena in born-infeld ads
  black holes}, \href{http://dx.doi.org/10.1103/PhysRevD.85.044040}{\emph{Phys.
  Rev. D} {\bf 85} (Feb, 2012) 044040}.

\bibitem{Liu1}
H.~Liu and X.-H. Meng, \emph{${P}-{V}$ criticality in the extended phase space
  of charged accelerating ads black holes},
  \href{http://dx.doi.org/10.1142/S0217732316501996}{\emph{Modern Physics
  Letters A} {\bf 31} (2016) 1650199},
  [\href{http://arxiv.org/abs/http://www.worldscientific.com/doi/pdf/10.1142/S0217732316501996}{{\tt
  http://www.worldscientific.com/doi/pdf/10.1142/S0217732316501996}}].

\bibitem{Gunasekaran1}
S.~Gunasekaran, D.~Kubiz{\v{n}}{\'a}k and R.~B. Mann, \emph{Extended phase
  space thermodynamics for charged and rotating black holes and born-infeld
  vacuum polarization},
  \href{http://dx.doi.org/10.1007/JHEP11(2012)110}{\emph{Journal of High Energy
  Physics} {\bf 2012} (2012) 1--43}.

\bibitem{Hendi1}
S.~H. Hendi, R.~M. Tad, Z.~Armanfard and M.~S. Talezadeh, \emph{Extended phase
  space thermodynamics and p-v criticality: Brans--dicke--born--infeld vs
  einstein--born--infeld-dilaton black holes},
  \href{http://dx.doi.org/10.1140/epjc/s10052-016-4106-9}{\emph{The European
  Physical Journal C} {\bf 76} (2016) 1--15}.

\bibitem{Zou1}
D.-C. Zou, S.-J. Zhang and B.~Wang, \emph{Critical behavior of born-infeld ads
  black holes in the extended phase space thermodynamics},
  \href{http://dx.doi.org/10.1103/PhysRevD.89.044002}{\emph{Phys. Rev. D} {\bf
  89} (Feb, 2014) 044002}.

\bibitem{Zou2}
D.-C. Zou, Y.~Liu and B.~Wang, \emph{Critical behavior of charged
  gauss-bonnet-ads black holes in the grand canonical ensemble},
  \href{http://dx.doi.org/10.1103/PhysRevD.90.044063}{\emph{Phys. Rev. D} {\bf
  90} (Aug, 2014) 044063}.

\bibitem{Mirza1}
B.~Mirza and Z.~Sherkatghanad, \emph{Phase transitions of hairy black holes in
  massive gravity and thermodynamic behavior of charged ads black holes in an
  extended phase space},
  \href{http://dx.doi.org/10.1103/PhysRevD.90.084006}{\emph{Phys. Rev. D} {\bf
  90} (Oct, 2014) 084006}.

\bibitem{Kastor1}
D.~Kastor, S.~Ray and J.~Traschen, \emph{Enthalpy and the mechanics of ads
  black holes}, {\emph{Class. Quantum Gravity} {\bf 26} (2009) 195011}.

\bibitem{Dolan1}
B.~Dolan, \emph{The cosmological constant and the black hole equation of
  state}, {\emph{Class. Quantum Gravity} {\bf 28} (2011) 125020}.

\bibitem{Dolan2}
B.~Dolan, \emph{Pressure and volume in the first law of black hole
  thermodynamics}, {\emph{Class. Quantum Gravity} {\bf 28} (2011) 235017}.

\bibitem{Dolan3}
B.~Dolan, \emph{Compressibility of rotating black holes}, {\emph{Phys. Rev. D}
  {\bf 84} (2011) 127503}.

\bibitem{Cvetic1}
M.~Cvetic, G.~Gibbons, D.~Kubiznak and C.~Pope, \emph{Black hole enthalpy and
  an entropy inequality for the thermodynamic volume}, {\emph{Phys. Rev. D}
  {\bf 84} (2011) 024037}.

\bibitem{Belhaj1}
A.~Belhaj, M.~Chabab, H.~E. Moumni and M.~Sedra, \emph{On thermodynamics of ads
  black holes in arbitrary dimensions}, {\emph{Chin. Phys. Lett.} {\bf 29}
  (2012) 100401}.

\bibitem{Hendi2}
S.~Hendi and M.~Vahidinia, \emph{Extended phase space thermodynamics and
  p–vcriticality of black holes with nonlinear source}, {\emph{Phys. Rev. D}
  {\bf 88} (2013) 084045}.

\bibitem{Spallucci1}
E.~Spallucci and A.~Smailagic, \emph{Maxwell’s equal area law for charged
  anti-de sitterblack holes}, {\emph{Phys. Lett. B} {\bf 723} (2013) 436--441}.

\bibitem{Altamirano1}
N.~Altamirano, D.~Kubiznak, R.~Mann and Z.~Sherkatghanad, \emph{Thermodynamics
  of rotating black holes and black rings: phase transitions and thermodynamic
  volume}, {\emph{Galaxies} {\bf 2} (2014) 89--159}.

\bibitem{Hendi3}
S.~H. Hendi, R.~B. Mann, S.~Panahiyan and B.~E. Panah, \emph{van der waals like
  behaviour of topological ads black holes in massive gravity}, {\emph{Phys.
  Rev. D} {\bf 95} (2017) 021501}.

\bibitem{Hendi4}
S.~H. Hendi, G.-Q. Li, J.-X. Mo, S.~Panahiyan and B.~E. Panah, \emph{New
  perspective for black hole thermodynamics in gauss-bonnet-born-infeld massive
  gravity}, {\emph{Eur. Phys. J. C} {\bf 76} (2016) 571}.

\bibitem{Hendi5}
S.~H. Hendi, S.~Panahiyan, B.~E. Panah, M.~Faizal and M.~Momennia,
  \emph{Critical behavior of charged black holes in gauss-bonnet gravity`s
  rainbow}, {\emph{Phys. Rev. D} {\bf 94} (2016) }.

\bibitem{Hendi6}
S.~H. Hendi, S.~Panahiyan and B.~E. Panah, \emph{Extended phase space of black
  holes in lovelock gravity with nonlinear electrodynamics}, {\emph{Prog.
  Theor. Exp. Phys.} {\bf 2015} (2015) 103E01}.

\bibitem{Johnson1}
C.~V. Johnson, \emph{Holographic heat engines}, {\emph{Class. Quant. Grav.}
  {\bf 31} (2014) 205002}.

\bibitem{Mo1}
J.-X. Mo, F.~Liang and G.-Q. Li, \emph{Heat engine in the three-dimensional
  spacetime}, \href{http://dx.doi.org/10.1007/JHEP03(2017)010}{\emph{Journal of
  High Energy Physics} {\bf 2017} (2017) 10}.

\bibitem{Johnson2}
C.~V. Johnson, \emph{Gauss-bonnet black holes and holographic heat engines
  beyond large n}, {\emph{Class. Quant. Grav.} {\bf 33} (2016) 215009}.

\bibitem{Johnson3}
C.~V. Johnson, \emph{Born-infeld ads black holes as heat engines},
  {\emph{Class. Quant. Grav.} {\bf 33} (2016) 135001}.

\bibitem{Belhaj2}
A.~Belhaj, M.~Chabab, H.~E. Moumni, K.~Masmar, M.~B. Sedra and A.~Segui,
  \emph{On heat properties of ads black holes in higher dimensions},
  {\emph{JHEP} {\bf 05} (2015) 149}.

\bibitem{Setare1}
M.~R. Setare and H.~Adami, \emph{Polytropic black hole as a heat engine},
  {\emph{Gen. Rel. Grav.} {\bf 47} (2015) 133}.

\bibitem{Caceres1}
E.~Caceres, P.~H. Nguyen and J.~F. Pedraza, \emph{Holographic entanglement
  entropy and the extended phase structure of stu black holes}, {\emph{JHEP}
  {\bf 1509} (2015) 184}.

\bibitem{Johnson4}
C.~V. Johnson, \emph{An exact efficiency formula for holographic heat engines},
  {\emph{Entropy} {\bf 18} (2016) 120}.

\bibitem{Wei1}
S.~W. Wei and Y.~X. Liu, \emph{Implementing black hole as efficient power
  plant}, {\emph{arXiv:1605.04629} (2016) }.

\bibitem{Zhang1}
M.~Zhang and W.-B. Liu, \emph{f(r) black holes as heat engines}, {\emph{Int J
  Theor Phys} {\bf 55} (2016) 5136--5145}.

\bibitem{Sadeghi1}
J.~Sadeghi and K.~Jafarzade, \emph{Heat engine of black holes},
  {\emph{arXiv:1504.07744} }.

\bibitem{Sadeghi2}
J.~Sadeghi and K.~Jafarzade, \emph{The correction of hoˇrava-lifshitz black
  hole from holographic engine}, {\emph{arXiv:1604.02973} }.

\bibitem{Bhamidipati1}
C.~Bhamidipati and P.~K. Yerra, \emph{Heat engines for dilatonic born-infeld
  black holes}, {\emph{arXiv:1606.03223} }.

\bibitem{Hennigar1}
R.~A. Hennigar, F.~McCarthy, A.~Ballon and R.~B. Mann, \emph{Holographic heat
  engines: general considerations and rotating black holes},
  {\emph{arXiv:1704.02314 [hep-th]} }.

\bibitem{Hendi7}
S.~H. Hendi, B.~E. Panah, S.~Panahiyan, H.~Liu and X.-H. Meng, \emph{Black
  holes in massive gravity as heat engines}, {\emph{arXiv:1707.02231 [hep-th]}
  }.

\bibitem{Mo2}
J.-X. Mo and G.-Q. Li, \emph{Holographic heat engine within the framework of
  massive gravity}, {\emph{arXiv:1707.01235 [gr-qc]} }.

\bibitem{Bachall1}
N.~Bachall, J.~Ostriker, S.~Perlmutter and P.~Steinhardt, \emph{The cosmic
  triangle: revealing the state of the universe}, {\emph{Science} {\bf 284}
  (1999) 1481}.

\bibitem{Perlmutter1}
S.~Perlmutter and et~al, \emph{Measurements of ${\Omega}$ and ${\Lambda}$-term
  from high-redshift supernovae}, {\emph{Astrophys. J.} {\bf 517} (1999) 565}.

\bibitem{Sahni1}
V.~Sahni and A.~Starobinsky, \emph{The case for a positive cosmological
  $\lambda$-term}, {\emph{Int. J. Mod. Phys. D} {\bf 9} (2000) 373}.

\bibitem{Fujii1}
Y.~Fujii, \emph{Origin of the gravitational constant and particle masses in a
  scale-invariant scalar-tensor theory}, {\emph{Phys. Rev. D} {\bf 26} (1982)
  2580}.

\bibitem{Ford1}
L.~Ford, \emph{Cosmological constant damping by unstable scalar fields},
  {\emph{Phys. Rev.D} {\bf 35} (1987) 2339}.

\bibitem{Ratra1}
B.~Ratra and P.~Peebles, \emph{Cosmological consequences of a rolling
  homogeneous scalar field}, {\emph{Phys. Rev. D} {\bf 37} (1988) 3406}.

\bibitem{Wetterich1}
C.~Wetterich, \emph{Cosmology and the fate of dilatation symmetry},
  {\emph{Nucl. Phys. B} {\bf 302} (1988) 668}.

\bibitem{Kiselev1}
V.~Kiselev, \emph{Quintessence and black holes}, {\emph{Class. Quantum Gravity}
  {\bf 20} (2003) 1187--1198}.

\bibitem{Chen1}
S.~Chen and J.~Jing, \emph{Quasinormal modes of a black hole surrounded by
  quintessence}, {\emph{Class. Quantum Gravity} {\bf 22} (2005) 4651--4657}.

\bibitem{Chen2}
S.~Chen, B.~Wang and R.~Su, \emph{Hawking radiation in a d-dimensional static
  spherically symmetric black hole surrounded by quintessence}, {\emph{Phys.
  Rev. D} {\bf 77} (2008) 124011}.

\bibitem{Varghese1}
N.~Varghese and V.~Kuriakose, \emph{Massive charged scalar quasinormal modes of
  reissner–nordstrom black hole surrounded by quintessence}, {\emph{Gen.
  Relativ. Gravit.} {\bf 41} (2009) 1249--1257}.

\bibitem{Tharanath1}
R.~Tharanath and V.~Kuriakose, \emph{Phase transition, quasinormal modes and
  hawking radiation of schwarzschild black hole in quintessence field},
  {\emph{Mod. Phys. Lett. A} {\bf 29} (2014) 1450057}.

\bibitem{Ainou1}
M.~Ainou and M.~Rodrigues, \emph{geometrical and poincaré methods for charged
  black holes in presence of quintessence}, {\emph{J. High Energy Phys.} {\bf
  1309} (2013) 146}.

\bibitem{Wei2}
Y.~Wei and Z.~Chu, \emph{Thermodynamic properties of a {R}eissner-{N}ordstroem
  quintessence black hole}, {\emph{Chin. Phys. Lett.} {\bf 28} (2011) 100403}.

\bibitem{Saleh1}
M.~Saleh, B.~Bouetou and T.~Kofane, \emph{Quasinormal modes of gravitational
  perturbation around a reissner-nordstroem black hole surrounded by
  quintessence}, {\emph{Chin. Phys. Lett.} {\bf 26} (2009) 109802}.

\bibitem{Zeng2}
X.-X. Zeng, D.-Y. Chen and L.-F. Li, \emph{Holographic thermalization and
  gravitational collapse in a spacetime dominated by quintessence dark energy},
  \href{http://dx.doi.org/10.1103/PhysRevD.91.046005}{\emph{Phys. Rev. D} {\bf
  91} (Feb, 2015) 046005}.

\bibitem{Zeng1}
X.-X. Zeng and L.-F. Li, \emph{Van der waals phase transition in the framework
  of holography},
  \href{http://dx.doi.org/http://doi.org/10.1016/j.physletb.2016.11.017}{\emph{Physics
  Letters B} {\bf 764} (2017) 100 -- 108}.

\bibitem{Chen3}
S.~Chen, Q.~Pan and J.~Jing, \emph{Holographic superconductors in quintessence
  ads black hole}, {\emph{Class. Quantum Grav.} {\bf 30} (2013) 145001}.

\bibitem{Li1}
G.-Q. Li, \emph{Effects of dark energy on p–vcriticality of charged ads black
  holes}, {\emph{Physics Letters B} {\bf 735} (2014) 256--260}.

\bibitem{Johnson5}
C.~V.Johnson, \emph{Approaching the carnot limit at finite power: An exact
  solution}, {\emph{arXiv:1703.06119} }.

\end{thebibliography}\endgroup
\end{document}